\documentclass[lettersize,journal]{IEEEtran}
\usepackage{amsmath,amsfonts}
\usepackage{algorithm}
\usepackage{array}
\usepackage[noend]{algpseudocode}
\usepackage[caption=false,font=normalsize,labelfont=sf,textfont=sf]{subfig}
\usepackage{textcomp}
\usepackage{stfloats}
\usepackage{url}
\usepackage{verbatim}
\usepackage{graphicx}
\usepackage{cite}
\begin{document}

\title{Application Scheduling with Multiplexed Sensing of Monitoring Points in Multi-purpose IoT Wireless Sensor Networks}

\author{Mustafa~Can~Çavdar,
        Ibrahim~Korpeoglu,
        and~Özgür~Ulusoy
\IEEEcompsocitemizethanks{\IEEEcompsocthanksitem M. C. Çavdar, I. Korpeoglu and Ö. Ulusoy are with the Department of Computer Engineering, Bilkent University, Cankaya, Ankara, 06800, Turkey.\protect\\
E-mail: mustafa.cavdar@bilkent.edu.tr, korpe@cs.bilkent.edu.tr, oulusoy@cs.bilkent.edu.tr
}
}



\maketitle

\begin{abstract}
Wireless sensor networks (WSNs) have many applications and are an essential part of IoT systems. The primary functionality of a WSN is gathering data from specific points that are covered with sensor nodes and transmitting the collected data to remote units for further processing. In IoT use cases, a WSN infrastructure may need to be shared by many applications, which requires scheduling those applications to time-share the node and network resources. In this paper, we investigate the problem of application scheduling in WSN infrastructures. We focus on the scenarios where applications request a set of monitoring points to be sensed in the region a WSN spans and propose a shared-data approach utilizing multiplexed sensing of monitoring points requested by multiple applications, which reduces sensing and communication load on the network. We also propose a genetic algorithm called GABAS, and three greedy algorithms for scheduling applications onto a WSN infrastructure considering different criteria. We performed extensive simulation experiments to evaluate our algorithms and compare them to some standard scheduling methods. The results show that our proposed methods perform much better than the standard scheduling methods in terms of makespan, turnaround time, waiting time, and successful execution rate metrics. We also observed that our genetic algorithm is very effective in scheduling applications with respect to these metrics.

\end{abstract}


\section{Introduction}
\IEEEPARstart{W}{ireless} sensor networks (WSNs) have become the key components of Internet-of-Things and smart environments due to improvements in sensing technologies, wireless communications, and mobile computing. Wireless sensor networks are heterogeneous systems consisting of sensor nodes that can collect different types of data from the points within their sensing range. The collected data can be processed at sensor nodes or higher-level distributed or centralized units, like base stations or cloud data centers.

The application types of WSNs are very broad. Some domains include smart cities, smart houses, and some other intelligent systems that are used in daily life. Smart city management is one of the major areas for which WSN applications are very useful. Intelligent parking systems~\cite{lee2008intelligent} and noise monitoring in metropolitan areas~\cite{maisonneuve2009citizen} are two examples of the applications that a smart city can make use of. Other examples of WSN applications include disaster prevention systems, agriculture management, habitat monitoring, intelligent lighting control, and supply-chain monitoring~\cite{taruna2011application}. 

Previously, WSNs were task-specific. A WSN was designed, developed, and optimized to support a single application. Another application deployment was impossible; therefore, most WSN resources were underutilized. However, recently, WSNs are started to be designed in a way that they can support multiple applications, similar to other systems. For instance, to a single city-wide WSN infrastructure, various types of applications such as air quality monitoring, noise monitoring, and crime detection can be deployed. Another example would be a single building-wide WSN, which can be used for both structural health monitoring~\cite{nigam2020wireless} and fire disaster detection~\cite{wahyono2020new} at the same time. Various other applications can be run over such a WSN infrastructure, such as occupancy estimation and automatic air-conditioning control, without disturbing other applications.
 
For a WSN that can handle multiple applications, it is essential that the WSN is designed and operated in such a way that applications get good quality of service and application owners are well satisfied. Achieving 
this usually requires a centralized mechanism and related policies. Software-Defined Networking (SDN) provides a mechanism that allows managing a WSN from a centralized controller~\cite{wang2016network}. SDN also enables virtualization that allows sharing physical resources among multiple services, tasks, or applications. Therefore, SDN is an essential component of next-generation networks and Internet-of-Things~\cite{sezer2013we}. With the help of SDN, applications can be scheduled and placed onto IoT-integrated WSNs with centralized algorithms in an efficient and effective way.

In WSNs that support running multiple applications over the same physical network, different applications may want the same data type (e.g., temperature, image) to be collected from the same points in the monitored area. For instance, there may be two applications, one of which monitors traffic density at a point and the other one measures the average speed of the vehicles between two points. The data collection frequency of these two applications may not be equal to each other, i.e., measuring average speed requires more frequent data collection than monitoring traffic density; however, both applications require the same type of data. Therefore, we propose a shared-data approach with multiplexed sensing for running multiple applications on a WSN. Our approach enables applications to share data from common monitoring points in the most efficient way, reducing the sensing and communication load incurred on WSNs. Even though the processing requirement will not change, the network will have more sensing and communication resources available to admit and schedule more applications simultaneously. This will help reduce the total execution time of a set of applications and also the waiting time of the newly arriving applications.

In this paper, we focus on the management of a WSN operated by a single infrastructure provider. The network is available to application providers who want their applications to be admitted to the network for a certain amount of time. Admitted applications need some points monitored with specific data types, and the collected data from those points need to be processed in base stations and centralized units. While we focused on the application placement problem in our previous work~\cite{CAVDAR2022109302} in a similar network structure, in this work, we deal with the application scheduling problem. Since the applications will use the network's resources for a particular time, it is crucial to schedule the applications (i.e., arrange the order of the admission of the applications into the network) efficiently and effectively. One of the critical parameters to minimize is the total execution time (makespan) of the applications, but there are other metrics that can be important, like average waiting time, turnaround time, and rate of completing the applications before deadlines, if any.

We propose several algorithms for application scheduling in wireless sensor networks. We first propose a genetic algorithm called GABAS that effectively schedules applications onto a sensor network. GABAS reduces the total execution time of the applications by both assigning monitoring points TO sensor nodes and base stations and determining the admission order in the best possible way. We also propose three greedy algorithms, considering different criteria, which can be used when fast decisions are needed.

We conducted extensive simulation experiments to evaluate and compare our algorithms with well-known standard scheduling algorithms. Experimental results show that GABAS is very effective in finding an admission order for applications, reducing total execution time. It also performs very well in other metrics such as average turnaround time, average waiting time, and successful completion ratio. Proposed greedy algorithms, on the other hand, are very fast and effective compared to other standard scheduling algorithms.

The rest of the paper is organized as follows: Section II gives and discusses the related work in literature. Section III presents our network model and problem formulation. Section IV describes our approach and algorithms in detail, and Section V provides the results of our simulation experiments. Finally, Section VI concludes the paper.

\section{Related Work}

Scheduling problem exhibits itself in all types of computer systems and networks, where resources are limited and there are tasks, applications, or services that need to time-share those resources. The resources can be the processors of a computer, the sensing and communication units of a wireless sensor network, the physical servers and switches of a cloud data center, or the edge computing nodes of a fog network.

There are many scheduling algorithms proposed in the literature for processor and cloud scheduling~\cite{abualigah2021novel, shukri2021enhanced, mirjalili2016multi, velliangiri2021hybrid, kang2021adaptive, sulaiman2021evolutionary, alboaneen2021metaheuristic, yang2020task, chen2020woa, chhabra2021performance, padhy2021mirage, sun2020contract, li2020load}. They use different meta-heuristics and evolutionary algorithms, such as particle swarm optimization, genetic algorithms, and ant colony optimization. Some of them also use greedy approaches or classical approaches for scheduling, such as first come first served, and shortest job first algorithms. The metrics they usually use include makespan, average response time, energy efficiency, utilization, execution cost, and average running time. They schedule tasks, jobs, or virtual machines on local or cloud computing components, like physical servers.  

There are also studies on scheduling in WSNs, IoT, fog, and edge computing. Fog and edge computing are usually integrated with WSNs in IoT systems. Porta et al.~\cite{porta2014sensor} propose EN-MASSE, a framework that deals with dynamic mission assignment for WSNs whose sensor nodes have energy harvesting capabilities. It is an integer programming method that assigns missions to sensor nodes and aims to minimize the total run-time of the missions. Uchiteleva et al.~\cite{uchiteleva2016virtualization} describe a resource scheduling algorithm for WSNs. The proposed scheduling algorithm is a resource management solution for isolated profiles in WSNs, and the authors compare their algorithm with Round Robin and Proportionally Fair scheduling algorithms. Wei et al.~\cite{wei2019q} present a Q-learning algorithm called ISVM-Q for task scheduling in WSNs. It optimizes application performance and total energy consumption. De Frias et al.~\cite{de2013scheduling} propose an application scheduling algorithm for shared actuator and sensor networks. Their algorithm aims to reduce energy consumption in the network. Edalat and Motani~\cite{edalat2016energy} propose a method for task scheduling and task mapping in a WSN consisting of sensor nodes with energy harvesting capabilities. They consider task priority and energy harvesting to increase fairness.

Liu et al.~\cite{liu2021task} propose Horae, a task scheduler for mobile edge computing. The scheduler aims to improve resource utilization in the MEC environment as well as select the edge server that satisfies placement constraints for each task. Javanmardi et al.~\cite{javanmardi2021fupe} present FUPE, a security-aware task scheduler for IoT fog networks. It is a fuzzy-based multi-objective Particle Swarm Optimization algorithm. The authors show that it performs better than the other compared algorithms in terms of average response time and network utilization. Li and Han~\cite{li2020hybrid} describe an artificial bee colony algorithm (ABC) for task scheduling in the cloud. The proposed ABC algorithm is compared against several works from the literature. The evaluation metrics they consider are makespan, maximum device workload, and total device workload. D'Amico and Gonzalez~\cite{d2021energy} propose EAMC, which is a multi-cluster scheduling policy. It predicts the energy consumption of jobs and aims to reduce makespan, response time, and total energy consumption. Singhal and Sharma~\cite{singhal2021job} present a Rock Hyrax Optimization algorithm to schedule jobs in heterogeneous cloud systems. They consider evaluation metrics like makespan and energy consumption. 

Choudhari et al.~\cite{choudhari2018prioritized} propose a  priority-based task scheduling algorithm for fog computing systems. Their algorithm first assigns an arriving request to the closest fog server, placing the task into a priority queue within that fog server. Xu et al.~\cite{xu2021optimal} apply online convex optimization techniques to schedule arriving jobs with multi-dimensional requirements in heterogeneous computing clusters. Psychasand and Ghaderi~\cite{psychasand2020high} describe algorithms based on Best Fit and Universal Partitioning to schedule jobs with various resource demands. Fang et al.~\cite{fang2019job} aim to reduce total job completion time in edge computing systems. They propose an approximation algorithm for both offline and online scheduling. Arri and Singh~\cite{arri2021energy} describe an artificial bee colony algorithm that also makes use of an artificial neural network for job scheduling in fog servers. 

Our work in this paper differs from the studies mentioned above with the following novel features:

\begin{itemize}
    \item We propose GABAS, a novel genetic algorithm that can schedule applications onto a WSN in an effective manner. While admitting and scheduling an application, our algorithm also decides which sensor and base stations will be used to sense and process data from the monitoring points requested by the application. Our genetic algorithm performs very well in various metrics, such as makespan, average turnaround time, average waiting time, and successful completion ratio.
    \item We also propose three greedy algorithms, each of which considers different criteria for ordering applications that are feasible to admit. These algorithms are better suited for scenarios where fast decisions are needed.
    \item We consider a network structure where certain points are monitored by sensor nodes in a multiplexed manner so that data can be shared and utilized by several applications at the same time. The sensing rate is adjusted depending on the demands of the applications requiring the common points to be sensed. The collected data is processed in base stations, and also in data centers if needed. In this way, sensing and communication resources of a WSN are better utilized, allowing more applications to be scheduled at the same time. We focus on urban-area networks, where base stations can easily be inter-connected with high-speed wired or wireless networks.
\end{itemize}

\section{Problem Statement}

The sensor network type we consider in this paper is a wireless sensor network (WSN) that is owned by a single provider and covers an urban region (like a city or town). Application owners require some points in the region to be sensed and sensed data to be collected and processed in cluster-heads. We assume the sensor network has a clustered two-tier architecture, consisting of sensor nodes and cluster-head nodes. We will call the cluster-head nodes also as base stations throughout the paper since they will act like base stations in a wireless network. They will be mains-powered and connected to the wired or wireless backbone network of a city. They will also be able to do local processing for the incoming sensor data. In an urban environment, cluster-head nodes can easily be mounted on top of the elements of the city-wide infrastructure (like lamp poles); therefore, they can easily be connected to the backbone network. In this way, the use of cluster-heads in our architecture is different from the classical WSN architectures, where cluster-heads are just intermediate nodes on multi-hop paths from sensor nodes to sink nodes. In an urban environment, we assume that each sensor node can directly connect to one of the base stations in its range. Each base station and the connected sensor nodes form a star topology.

We assume sensor nodes have equal sensing rates but various sensing ranges. Base stations have equal processing capacity. Additionally, we assume that any link between a sensor node and a base station has the same bandwidth capacity. A sensor node can collect data from the monitoring points which fall into its sensing range. The collected data is processed, partially or totally, in the base stations. It is possible that a base station can send the processed data further to a centralized location for additional processing or analysis. However, this part of the problem is not within the scope of this paper. We assume the bandwidth of the links connecting base stations to the rest of the network is abundant; hence is not a constraint in our formulation. We only consider the bandwidth capacity of connections between sensor nodes and base stations as a constraint. Similarly, we assume that the centralized servers have the abundant capacity to process the incoming data if needed. Therefore, we only consider the processing capacity of the base stations as a constraint in our model.

\begin{figure}
    \centering
     \includegraphics[width=\linewidth]{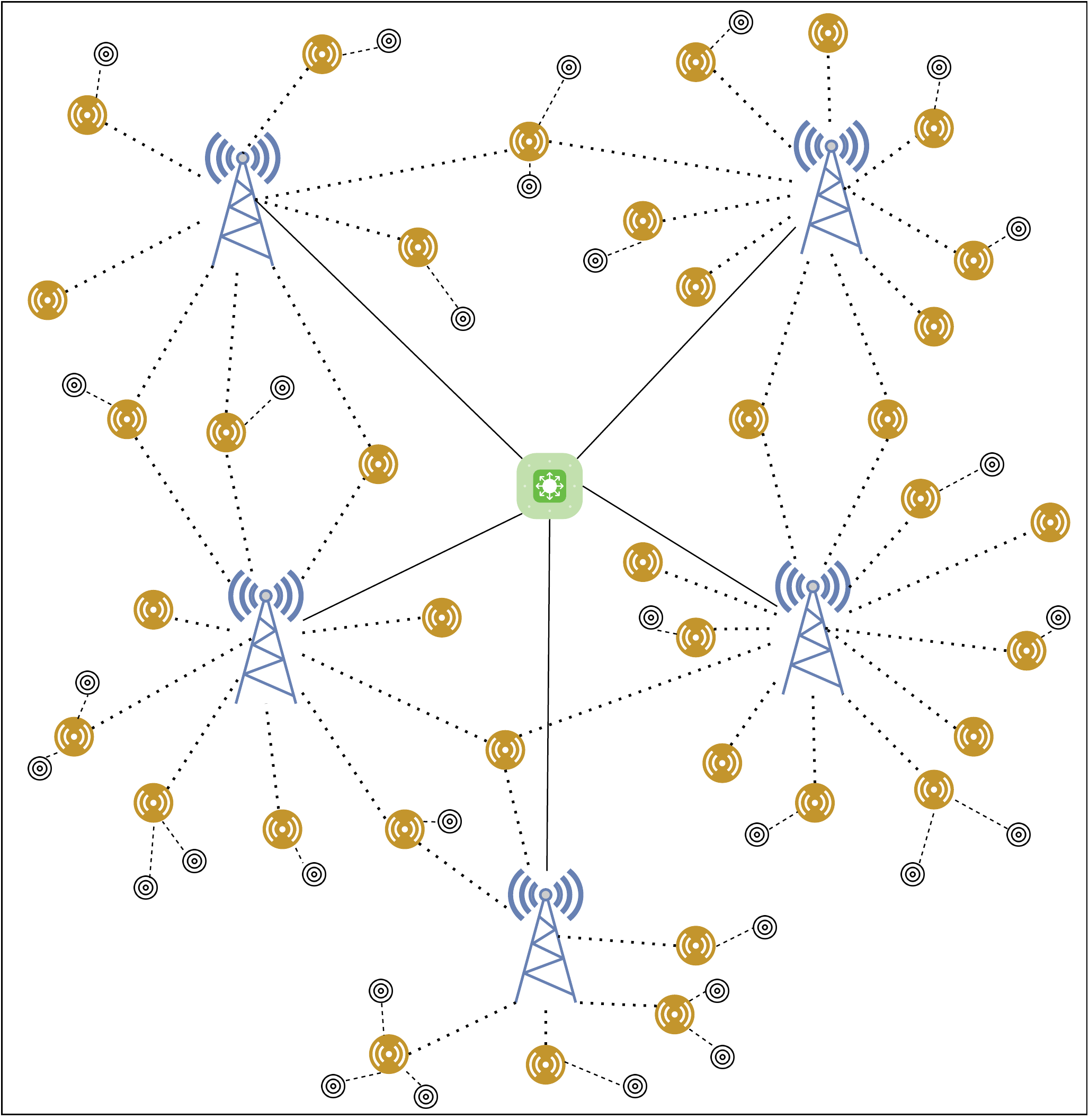}
    \caption{Network model.}
    \label{networkschema}
\end{figure}

Figure~\ref{networkschema} shows an example of our network model. In the figure, radio towers represent base stations (cluster-heads), large circles represent sensor nodes and small circles represent monitoring points. A dashed line between a sensor node and a monitoring point indicates that the sensor node actively senses the monitoring point. If a sensor node is connected to a base station, there is a dotted line between the sensor node and the base station. Base stations are connected to each other through a high-speed wired or wireless network.

Applications that arrive at the network require some points to be sensed and related data to be processed for a certain amount of time. One important goal is to finish all applications using the network as early as possible. If the requirements of an application are satisfiable and the application is scheduled to use the network, the application is deployed to the network and uses the required resources for the desired amount of time. After this period of time has expired, the application releases the resources and leaves the network. 

The parameters used for a formal description of our problem are shown in Table \ref{params}.
\newline

\begin{table}
\centering
\caption{Parameters used in the problem statement.}
\label{params}
\begin{tabular}{|l|c|}
\hline
$S$ & Set of sensor nodes \\ \hline
$B$ & Set of base stations \\ \hline
$C$ & Set of connections \\ \hline
$A$ & Set of applications \\ \hline
$M$ & Set of monitoring points \\ \hline
$T$ & Set of time instants \\ \hline
$t_{j}$ & Time period that application \textit{j} wants to use the network \\ \hline
$t0_{j}$ & Time at which application \textit{j} is admitted \\ \hline
$tf_{j}$ & Time at which application \textit{j} finishes \\ \hline
$tf_{max}$ & Time the last application finishes \\ \hline
$x_{jt}$ & \begin{tabular}{c} Binary variable indicating whether application \textit{j}\\ is deployed at time \textit{t} \end{tabular}  \\ \hline
$M_{j}$ & \begin{tabular}{c} Set of monitoring points to be sensed\\ for application \textit{j} \end{tabular}  \\ \hline
$r_{jk}$ & \begin{tabular}{c}Sensing rate requirement of application \textit{j} \\
for monitoring point \textit{k} \end{tabular}  \\ \hline
$r_{kt}$ & Sensing rate requirement of monitoring point \textit{k} at time \textit{t}\\ \hline
$S_{k}$ & Set of sensor nodes covering monitoring point \textit{k}  \\ \hline
$M_{il}$ & \begin{tabular}{c}  Set of monitoring points whose data is transferred\\ from sensor node \textit{i} to base station \textit{l} \end{tabular} \\ \hline
$x_{ik}$ & \begin{tabular}{c}Binary variable indicating whether sensor node \textit{i} \\is actively sensing monitoring point \textit{k} \end{tabular} \\ \hline
$x_{ilm}$ & \begin{tabular}{c}Binary variable indicating whether sensor node \textit{i} is \\connected to base station \textit{l} through connection \textit{m}\end{tabular}\\ \hline

$r_k$ & Sensing rate requirement of monitoring point \textit{k} \\ \hline
$R_{i}$ & Sensing rate capacity of sensor node \textit{i}\\ \hline
$P_{l}$ & Processing capacity of base station \textit{l}\\ \hline
$C_{m}$ & Bandwidth capacity of connection \textit{m}\\ \hline
$\alpha$ & Transmission coefficient \\ \hline
$\beta$ & Processing coefficient \\ \hline
\end{tabular}
\end{table}

Our primary goal is to minimize the total run-time of applications requiring WSN service. This optimization problem can be formalized as follows:

\begin{equation}\label{time}
tf_{max}
\end{equation}

is minimized subject to

\begin{equation}\label{maxtime}
   tf_{max} = max(tf_{j}) \quad \quad \forall j \in A
\end{equation}

\begin{equation}\label{finishtime}
   tf_{j} = t0_{j} + t_{j} \quad \quad \forall j \in A
\end{equation}

\begin{equation}\label{apptime}
   x_{jt} = 
   \begin{cases}
   1, & t0_{j} < t \leq tf_{j} \\
   0, & otherwise
   \end{cases}
   \quad \quad \forall j \in A
\end{equation}

\begin{equation}\label{mpreqtime}
   r_{kt} = max(r_{jk} \times x_{jt}) \quad \quad \forall j \in A, \forall t \in T, \forall k \in M
\end{equation}

\begin{equation}\label{mpusrtime}
    u_{kt} = \sum_{j \in A} (r_{jk} \times x_{jt}) \quad \quad \forall k \in M, \forall t \in T
\end{equation}

\begin{equation}\label{senseconstrtime}
    \sum_{k \in M} x_{ik}r_{kt} \leq R_i \quad \quad \forall i \in S, \forall t \in T
\end{equation}

\begin{equation}\label{connconstrtime}
    \sum_{k \in M_{il}} \alpha r_{kt} \leq \sum_{m \in C} x_{ilm}C_{m} \quad  \forall i \in S, \forall l \in B, \forall t \in T
\end{equation}

\begin{equation}\label{procconstrtime}
    \sum_{i \in S}\sum_{k \in M_{il}} \beta u_{kt} \leq P_{l} \quad \quad \forall l \in B, \forall t \in T
\end{equation}

Eq. \ref{maxtime} shows how total run-time (makespan) is calculated. It is the finish time of the last application that used the network. We are assuming the first application is admitted at time 0. Eq.\ref{finishtime} describes how the finish time of each application is determined. Eq. \ref{apptime} is used to determine the time interval an application is deployed to the network. Eqs. \ref{mpreqtime} and \ref{mpusrtime} show the calculation of sensing requirements (i.e., required sensing rates) of monitoring points in shared and unshared cases, respectively. Eqs. \ref{senseconstrtime}, \ref{connconstrtime}, and \ref{procconstrtime} are sensing, transmission and processing constraints for each time instant. At each time instant, the total sensing rate requirement of monitoring points sensed by one sensor node must not exceed that sensor node's sensing capacity. The data amount transmitted per second cannot be more than the bandwidth of that connection. Also, the total processed data per second at one base station should be less than the base station's processing capacity. $\alpha$ (transmission coefficient) is a constant that maps a sensing rate value to a communication rate requirement. Similarly, $\beta$ (processing coefficient) is a constant used to map a sensing rate value to a processing capacity requirement. For example, if data is sensed at a rate of 3 Kbps, the required communication rate may be 4 Kbps, including communication protocol overheads.

\subsection{Hardness of Application Scheduling}

The application scheduling problem explained above is a variation of the well-known task scheduling problem. The task scheduling problem is proven to be an NP-hard problem~\cite{karp1972reducibility}. To prove that application scheduling is also an NP-hard problem, we reduce the multiway number partitioning problem to it. Reduction is shown in Appendix A.

\section{Proposed Methods}

\subsection{Genetic Algorithm Based Application Scheduling (GABAS)}

\begin{figure}[ht]
    \centering
    \includegraphics[width=\linewidth]{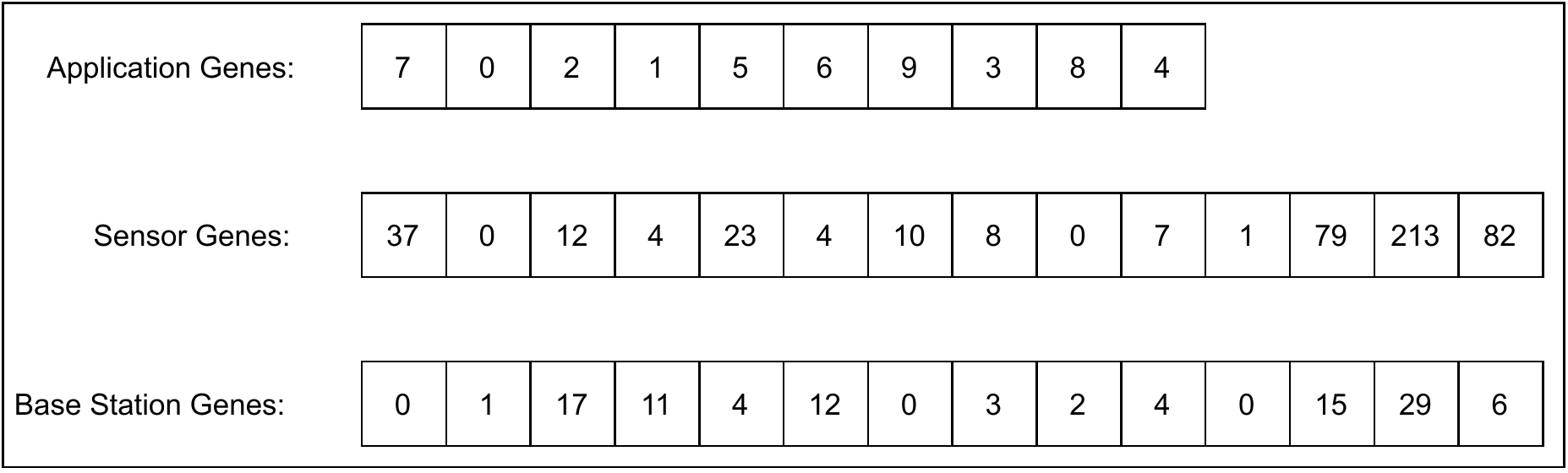}
    \caption{Example genes.}
    \label{lists}
\end{figure}

To effectively schedule applications to a wireless sensor network, we propose a novel genetic algorithm called GABAS. A genetic algorithm is a meta-heuristic solution that mimics natural selection. It runs for \textit{generations} until a termination condition is met. Each generation consists of \textit{individuals}, which are candidate solutions for the problem. The actual solution that an individual provides is called its \textit{chromosome}. The chromosome structure of an individual consists of one or more genes, each of which is a list or array, usually. In GABAS, the chromosome structure of individuals contains three lists of integers identifying applications, sensor nodes, and base stations. The first list is \textit{Application Genes} which represents the scheduling order of applications. Its size is equal to the number of applications. The other two lists are \textit{Sensor Genes} and \textit{Base Station Genes}, which represent the proposed sensor nodes and base stations for the monitoring points, respectively. The size of the lists is equal to the number of monitoring points in the whole network area. Examples of these lists are shown in Figure \ref{lists}.
 
\subsubsection{Initial Population Creation}

In genetic algorithms, individuals are created by crossover operation. Therefore, their \textit{genes} are determined by two individuals from the previous generation. However, since there is no previous generation for the first generation, we need to create the individuals of the first generation randomly.

Since \textit{Application Genes} of an individual determines the admission order of the applications, we assign value $i$ to the $i^{th}$ application gene and shuffle the list. Therefore, initially, the admission order is randomly determined.

For the individuals of the initial population, \textit{Sensor Genes} and \textit{Base Station Genes} are determined together. For each monitoring point, first, we randomly determine a sensor gene among the sensor nodes that cover the monitoring point. After that, we randomly select a base station gene among the base stations to which the selected sensor node has a connection.

\subsubsection{Fitness Calculation}

Fitness calculation in a genetic algorithm is designed and used to measure how close an individual is to the optimum solution. The result of fitness calculation is called \textit{fitness value}. Equation~\ref{fitness} shows the fitness calculation of our algorithm.

\begin{equation}\label{fitness}
    fitness = -1 \times makespan
\end{equation}

Basically, GABAS aims to reduce the makespan, which is the time instant when the last application finishes and leaves the network. Since a higher fitness value means a better individual (solution), we multiply the makespan with $-1$, since a lower makespan value is a more desirable one. 

Applications are admitted one by one according to their order in Application Genes. If an application cannot be admitted due to a shortage of resources, we wait for some already admitted applications to finish and release the resources they use. Then there will be sufficient resources available for the waiting application. 

\subsubsection{Selection Operation}

The selection operation is performed to choose an individual to pair up for each individual in the population of the current generation to create the next generation. We use tournament selection for this process. Basically, for each individual $i$, we create a sub-population consisting of 5\% of the whole population and select the individual with the highest fitness score in this subpopulation. Then we pair the individual $i$ with the selected individual. Our selection operation is presented in Algorithm \ref{selection}.

\begin{algorithm}[]
\caption{Selection Operation.}
\label{selection}
\begin{algorithmic}[1]
\Require{The tournament population, $tPop$}
\Ensure{an individual chromosome}
\Procedure{Selection}{}
\State{$\textit{$bestScore$} \leftarrow 0$}
\State{$\textit{$bestInd$} \leftarrow Null$}
\For{each Individual $x$ in $tPop$}
\State{calculate its fitness score, $newScore$}
\If{$bestScore \leq newScore$}
\State{$\textit{$bestScore$} \leftarrow newScore$}
\State{$\textit{$bestInd$} \leftarrow x$}
\EndIf
\EndFor

\Return $bestInd$
\EndProcedure
\end{algorithmic}
\end{algorithm}

\subsubsection{Crossover Operation}

The crossover operation is executed to create the population of the next generation. Individuals that are paired up with the selection operation are used in the crossover operation. The operation determines which genes of the offspring come from which parent (a pair of individuals). This operation is presented in Algorithm \ref{crossover}. We assume the chance of either parent to pass its genes to the offspring is 50\%; therefore, we have \textit{uniformRate} value in the algorithm as 0.5. The Crossover operation of \textit{Sensor Genes} and \textit{Base Station Genes} is realized together to avoid producing a candidate solution that conflicts with the network structure. 

Crossover of the \textit{Application Genes} may result in some applications appearing twice in the Application Genes of the offspring. Therefore, we need a \textit{gene repairing} algorithm to fix this problem. For that, we first determine the applications that appear twice and those that do not appear at all in the offspring's genes. Then, we put applications that are missing into the places of second appearances of the applications that are present twice in the genes.

\begin{algorithm}[]
\caption{Crossover Operation.}
\label{crossover}
\begin{algorithmic}[1]
\Require{Six chromosomes from two parents: \textit{$A_{1}$}, \textit{$S_{1}$}, \textit{$BS_{1}$}, \textit{$A_{2}$}, \textit{$S_{2}$}}, and \textit{$BS_{2}$}
\Ensure{Three offspring chromosomes: \textit{$A_{new}$}, \textit{$S_{new}$}} and \textit{B$S_{new}$}
\Procedure{Crossover}{}
\For{$x = 1 \:\:to \:\:|\textit{\textbf{A}}|\:\: $}
\State{randomly create a value between 0 and 1, $r$;}
\If {$r \leq uniformRate$} 
\State{set gene $x$ of \textit{$A_{new}$} as gene $x$ of \textit{$A_{1}$}}
\Else
\State{set gene $x$ of \textit{$A_{new}$} as gene $x$ of \textit{$A_{2}$}}
\EndIf
\EndFor
\For {$x = 1 \:\:to \:\:|\textit{\textbf{M}}|\:\: $}
\State{randomly create a value between 0 and 1, $r$;}
\If {$r \leq uniformRate$} 
\State{set gene $x$ of \textit{$S_{new}$} as gene $x$ of \textit{$S_{1}$}}
\State{set gene $x$ of \textit{$BS_{new}$} as gene $x$ of \textit{$BS_{1}$}}
\Else
\State{set gene $x$ of \textit{$S_{new}$} as gene $x$ of \textit{$S_{2}$}}
\State{set gene $x$ of \textit{$BS_{new}$} as gene $x$ of \textit{$BS_{2}$}}
\EndIf
\EndFor
\Return A new individual with chromosomes: \textit{$A_{new}$}, \textit{$S_{new}$} and \textit{B$S_{new}$}
\EndProcedure
\end{algorithmic}
\end{algorithm}

\subsubsection{Mutation Operation}

The mutation operation is applied to all individuals. The operation is presented in Algorithm~\ref{mutation}. As in the crossover operation, mutation operation in \textit{Sensor Genes} and \textit{Base Station Genes} is realized together to guarantee a solution that abides the network structure. This operation is executed for each gene with a chance of 5\%.

For mutation in \textit{Application Genes}, we swap the admission order of the two randomly selected applications with a 5\% mutation rate. 

\begin{algorithm}[]
\caption{Mutation Operation.}
\label{mutation}
\begin{algorithmic}[1]
\Require{Three chromosomes: \textit{$A_{old}$}, \textit{$S_{old}$} and \textit{$BS_{old}$}}
\Ensure{Three mutated chromosomes: \textit{$A_{new}$}, \textit{$S_{new}$} and \textit{$BS_{new}$}}
\Procedure{Mutation}{}
\State{\textit{$A_{new}$} $\leftarrow$ \textit{$A_{old}$};}
\State{randomly create a value between 0 and 1, $r1$;}
\If {$r1 \leq mutationRate$}
\State{randomly create a value between 0 and $|M|$, $x$;}
\State{randomly create a value between 0 and $|M|$, $y$;}
\State{swap gene $x$ and gene $y$ of \textit{$A_{new}$}}
\EndIf
\For {$x = 1 \:\:to \:\:|\textit{\textbf{M}}|\:\: $}
\State{randomly create a value between 0 and 1, $r2$;}
\If {$r2 \leq mutationRate$}
\State{randomly select a sensor node \textit{i} from possible sensor nodes, where $1 \leq i \leq |S|$}
\State{randomly select a base station \textit{l} from possible base stations, where $1 \leq l \leq |B|$}
\State{set gene $x$ of \textit{$S_{new}$} as gene $i$}
\State{set gene $x$ of \textit{$BS_{new}$} as gene $l$}
\Else
\State{set gene $x$ of \textit{$S_{new}$} as gene $x$ of \textit{$S_{old}$}}
\State{set gene $x$ of \textit{$BS_{new}$} as gene $x$ of \textit{$BS_{old}$}}
\EndIf
\EndFor
\Return Mutated individual with chromosomes: \textit{$A_{new}$}, \textit{$S_{new}$} and \textit{B$S_{new}$}
\EndProcedure
\end{algorithmic}
\end{algorithm}

\subsubsection{The Genetic Algorithm}

Our overall genetic algorithm is presented in Algorithm \ref{thega}. The termination condition of the algorithm is that no improvement is observed in the fitness score of the best individual for seven generations. The population size is 200. The fitness calculation for each individual is $-1$ times the completion time of the last finished application. All individuals have negative fitness scores since a better solution means a shorter finish time. Elitism is enabled in the algorithm, which means that the fittest individual of one generation is carried over to the next generation.

\begin{algorithm}[]
\caption{The Genetic Algorithm.}
\label{thega}
\begin{algorithmic}[1]
\Procedure{The GA}{}

generate a population of $POPSIZE$ number of random individuals, $POP$;
\While{THE TERMINATION CONDITION is not $true$}
\For{each Individual $x$ in $POP$}
\State{calculate its fitness value $f(x)$}
\EndFor
\For{each Individual $x$ in $POP$}
\State Create a tournament population, $tPop$
\State{$y$ = SelectionOperation($tPop$)}
\EndFor
\For{each pair of parents, $x$ and $y$}
\State{$z$ = CrossoverOperation($x, y$)}
\EndFor
\For{each offspring}
\State{$z$ = MutationOperation($z$)}
\EndFor
\State{find the best individual among offsprings, $newBest$}
\If{$newBest$ is better than the current best individual}
\State{replace current best individual with $newBest$}
\EndIf
\EndWhile

\Return best individual
\EndProcedure
\end{algorithmic}
\end{algorithm}

Next, we describe our greedy algorithms for WSN application scheduling.

\subsection{Greedy Algorithms}

For scenarios where fast decisions are needed, we also propose three simple greedy algorithms for scheduling applications onto a sensor network. All these greedy algorithms use the \textit{Worst Fit} approach in assigning monitoring points to sensor nodes and base stations. The \textit{Worst Fit} approach uses the less utilized resources among the available ones to provide load balancing. The difference between the three algorithms is the criteria they use for ordering the applications to admit to the network.

Our greedy algorithms consider only the waiting applications to determine their order. The applications that are already admitted to the network are allowed to run until they are complete. We do not preempt a running application. 

Next, we give more information about each of our greedy algorithms.

\subsubsection{Least Monitoring Point First (LMPF)}

In LMPF algorithm, we order the applications according to the number of monitoring points they require to be sensed. Among waiting applications, the ones with a smaller number of monitoring points to be sensed are admitted earlier.

\subsubsection{Least Maximum Sense Requirement First (LMSF)}

In LMSF, the order of applications is determined according to their sensing rate requirements. Applications are admitted in an order in which they are sorted according to their maximum sensing rate requirement from a monitoring point (among all monitoring points requested to be sensed by an application). The order is a non-decreasing one; therefore, a waiting application with the least maximum requirement is placed first onto the network.

\subsubsection{Least Total Sense Requirement First (LTSF)}

In LTSF, the application admission order is again determined according to the sensing rate requirements. However, for an application, instead of considering the maximum sensing rate requirement from a monitoring point, we consider the sum of sensing rate requirements for all its monitoring points.

\section{Experimental Results}

The network model in our simulations spans a 2D plane of size 1000 m $\times$ 1000 m. Monitoring points, sensor nodes, and base stations have their coordinates (\textit{x} and \textit{y}), which are randomly determined and do not overlap. Therefore, at each coordinate, there is at most one network element.

We assume that applications arriving at the network may only require sensing rates on three major scales. Sense rate requirements of applications for each scale are randomly determined between the values shown in Table \ref{senserates}. We ensure that only one data type can be requested from a single monitoring point. Our network constraints are shown in Table \ref{networkparams}. Each application can request 1, 2, or 3 monitoring points to be sensed. The communication range of a sensor node is randomly selected between 200 m and 250 m, and the range determines to which base stations the sensor node can get connected. Similarly, the sensing range for each sensor node is between 30 m and 50 m, and it is used to determine which monitoring points can be sensed by the sensor node. Additionally, the number of sensor nodes and base stations is 250 and 30, respectively. We observe that these values are large enough to cover the whole network area and small enough to let us understand the performance differences of the compared algorithms. The number of applications is 1000, and the number of monitoring points is 300. Applications arrive at the network in 25 batches, and the batch number for each application is randomly determined.

\begin{table}[H]
\centering
\caption{Sensing rate requirements.}
\begin{tabular}{|l|c|}
\hline
\multicolumn{1}{|c|}{\textbf{Data Type}} & \textbf{sensing rate} \\ \hline
Data Type 0 & 5 - 20 \\ \hline
Data Type 1 & 15 - 40 \\ \hline
Data Type 2 & 25 - 60 \\ \hline
\end{tabular}
\label{senserates}
\end{table}

\begin{table}[H]
\centering
\caption{Network constraints.}
\begin{tabular}{|l|c|}
\hline
\multicolumn{1}{|c|}{\textbf{Constraint}} & \textbf{Value} \\ \hline
Monitoring Points per Application & 1 - 3 \\ \hline
Communication Range of Sensors & 200 - 250 \\ \hline
Sensing Range of Sensors & 30 - 50 \\ \hline
\end{tabular}
\label{networkparams}
\end{table}

Our simulation experiments are realized to investigate how certain parameters of the network affect the performance of algorithms that are compared. Unless otherwise stated below, the parameter values are set as explained above. For each scenario, we provide makespan, waiting time, turnaround time, and successful execution rate values averaged over 100 runs. We divide our experiments into six scenarios as follows:

\begin{itemize}
    \item \textit{Scenario 1 (Application Count)}: The number of applications starts at 500 and is incremented by 100 until 1500.
    \item \textit{Scenario 2 (Monitoring Point Count)}: We start with 50 monitoring points in the area and increment the total number of monitoring points by 25 until 250.
    \item \textit{Scenario 3 (Monitoring Point Count per Application)}: The number of monitoring points requested per application is initially 1. It is incremented by 1 until 7.
    \item \textit{Scenario 4 (Communication Range)}: In the beginning, each sensor node has a 50 m communication range. We increase the communication range by 50 m until 250 m.
    \item \textit{Scenario 5 (Sensing Range)}: Similar to Scenario 4, each sensor node starts with a 30 m sensing range, and the sensing range is incremented by 5 m until 50 m.
    \item \textit{Scenario 6 (Batch Count)}: We experiment with the following values for the number of batches (batch count): 1, 2, 5, 10, 20 and 25. Applications are equally distributed into batches. The number of applications in each batch (batch size) is the number of applications divided by the batch count.
    
\end{itemize}

We use four different metrics to provide a comparative evaluation of our proposed algorithms.
\begin{itemize}
    \item Average Makespan: Average of the total execution time of applications in 100 runs.
    \item Average Waiting Time: The waiting time of an application is the time between its arrival to the network and its admission. We report the average of the waiting times of all applications admitted.
    \item Average Turnaround Time: The turnaround time of an application is the time between its arrival time to the network and its finish time. We report the turnaround time averaged over all applications admitted.
    \item Average Successful Execution Rate: This metric is the number of applications that could finish before their deadlines. For each application, a deadline value is determined. The deadline value for an application is set to be the sum of its arrival time, its required running time (application duration), and a random value between 100 and 200. 
\end{itemize}

We compare our proposed algorithms with well-known standard task scheduling algorithms First Come First Served (FCFS) and Shortest Job First (SJF). For FCFS and SJF algorithms, we also used the \textit{Worst Fit} approach to determine which sensor node and the base station are assigned to each monitoring point. In the figures provided below, both shared (GABAS-S) and unshared (GABAS-U) approaches for GABAS are reported. For other algorithms, only shared approach results are provided to improve the readability since their performance with the shared approach is always better than the unshared approach.

\begin{figure}[]
    \centering
    \includegraphics[width=0.45\textwidth]{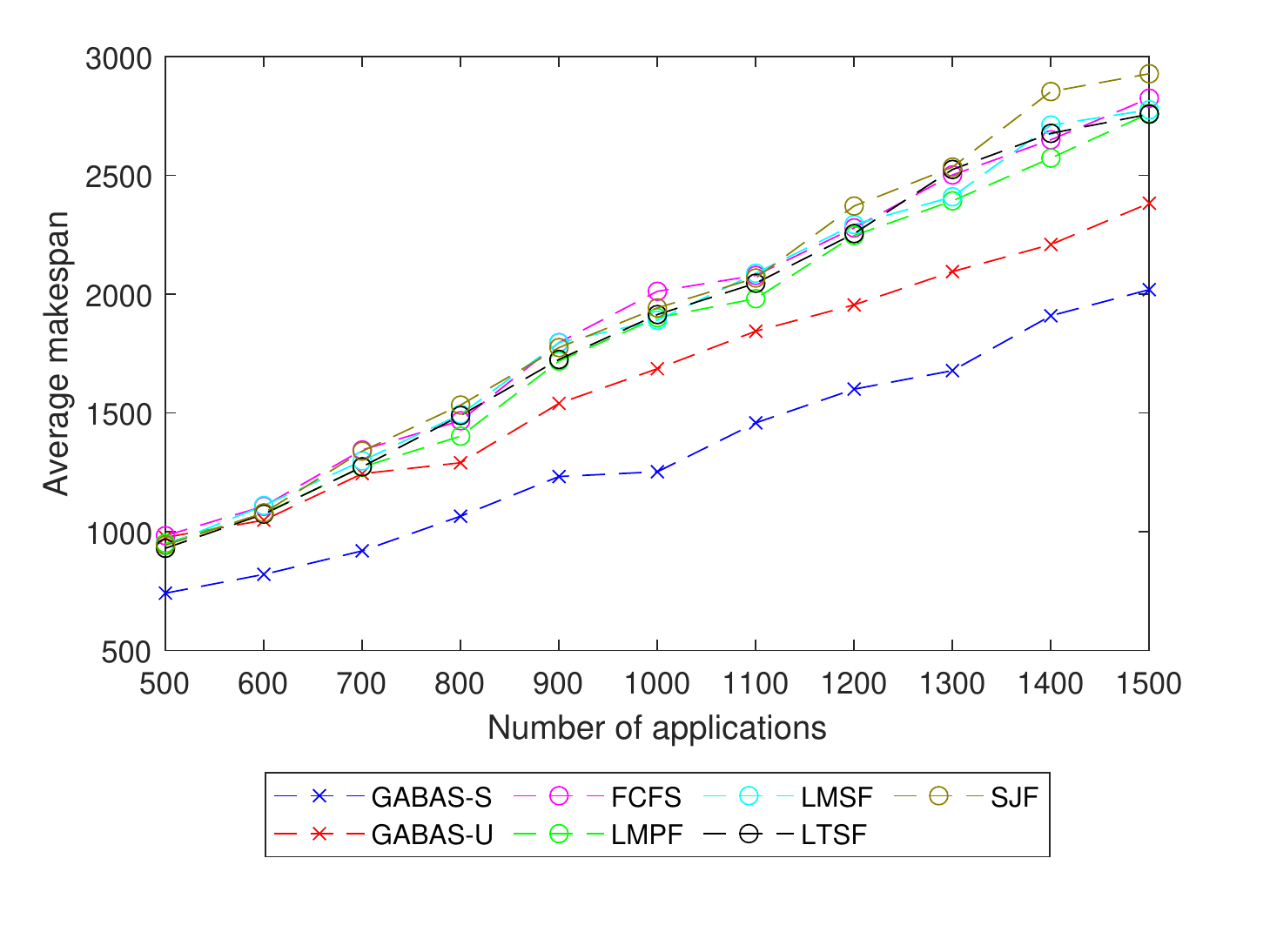}
    \caption{Comparison of algorithms in terms of average makespan in Scenario 1.}
    \label{scenario1avg}
\end{figure}

In Scenario 1, we investigate how the number of applications affects the performance of the algorithms. Figure \ref{scenario1avg} shows the average makespan. GABAS-S has the best performance compared to others. GABAS-U is the worst for the small number of applications. However, with the increasing number of applications, its performance becomes better than the greedy ones. Our LMPF algorithm comes third, while LMSF and LTSF have similar performance as FCFS and SJF.

\begin{figure}[]
    \centering
    \includegraphics[width=0.45\textwidth]{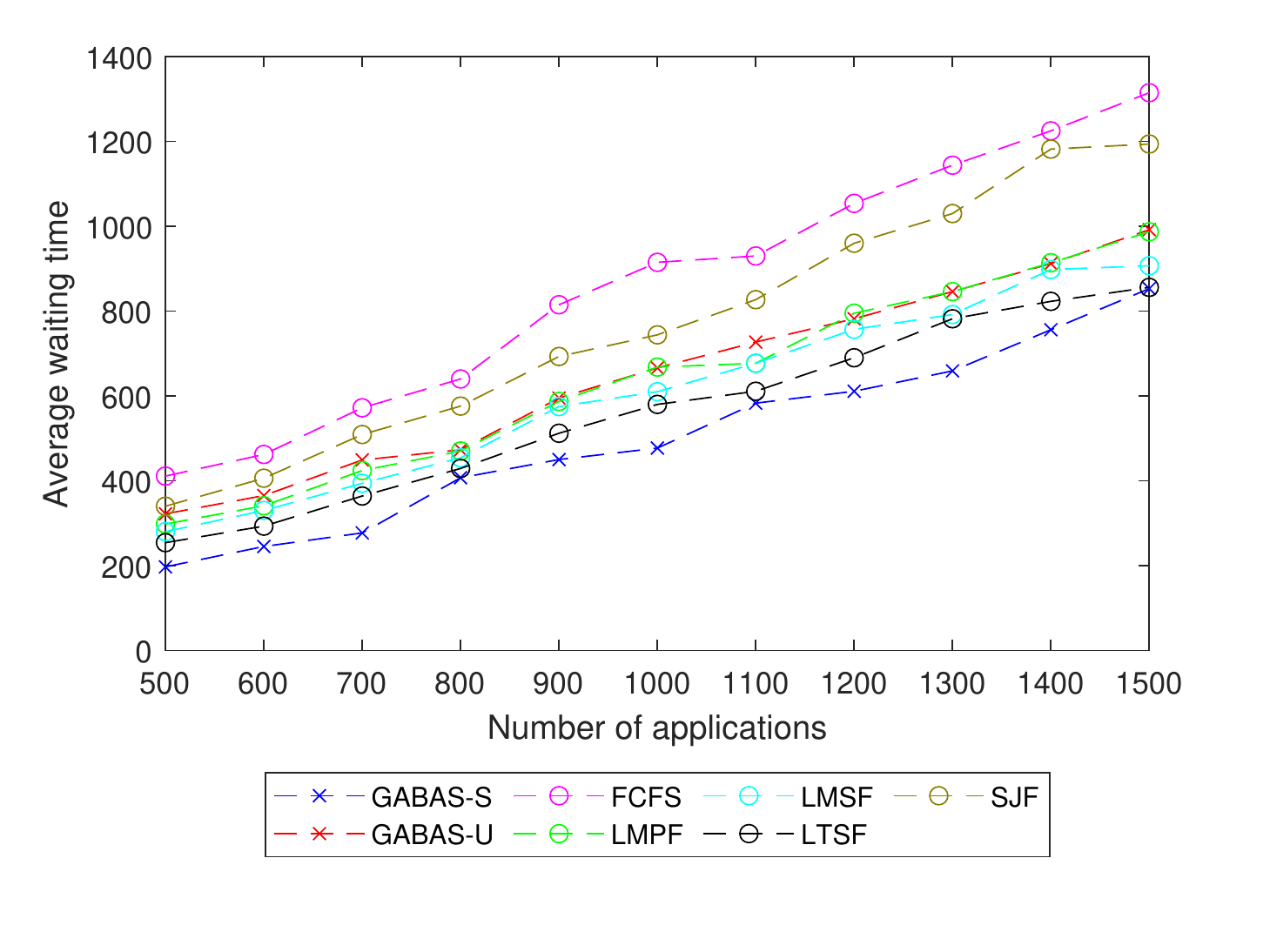}
    \caption{Comparison of algorithms in terms of average waiting time in Scenario 1.}
    \label{scenario1wai}
\end{figure}

\begin{figure}[]
    \centering
    \includegraphics[width=0.45\textwidth]{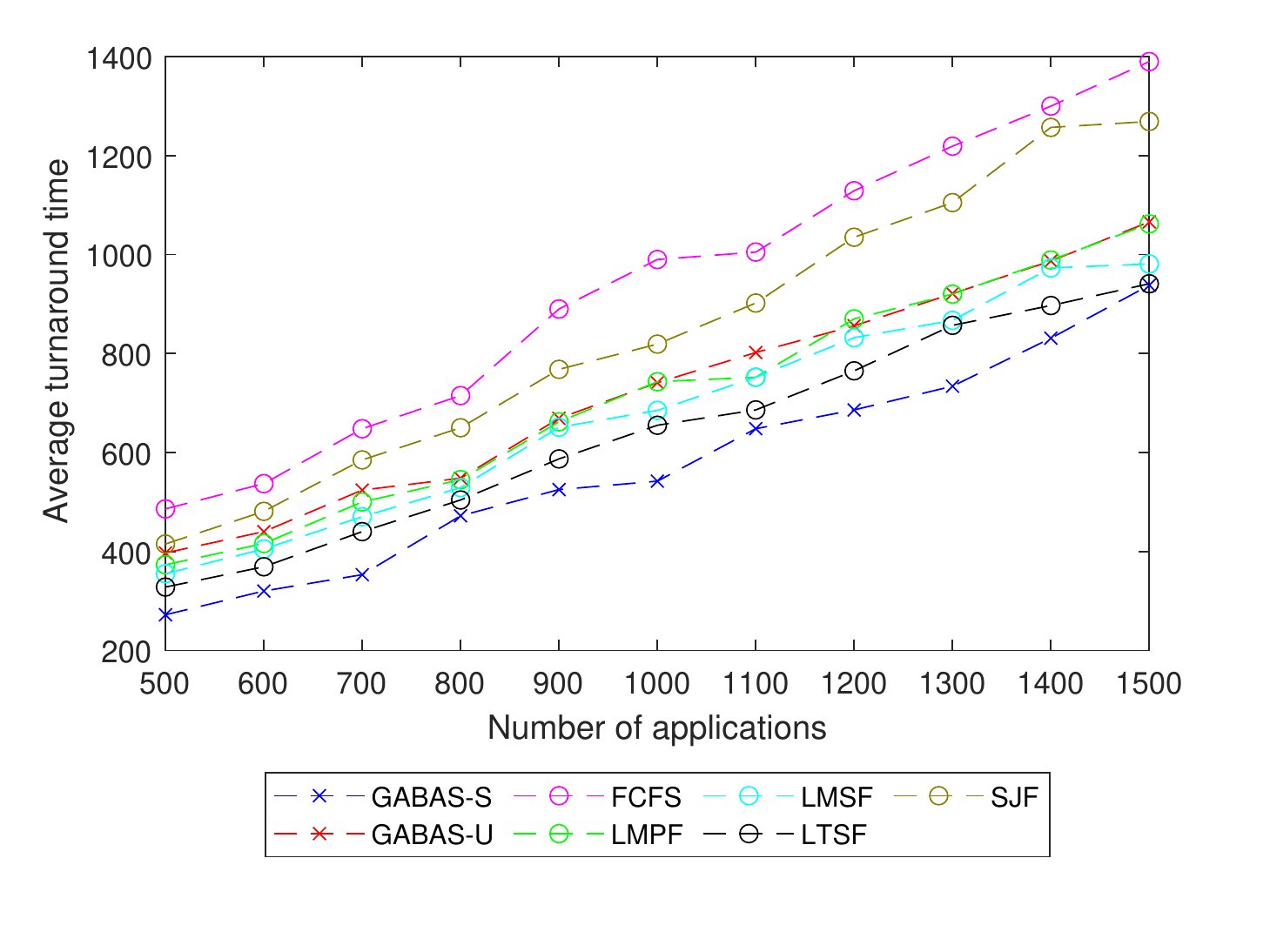}
    \caption{Comparison of algorithms in terms of average turnaround time in Scenario 1.}
    \label{scenario1res}
\end{figure}

\begin{figure}[]
    \centering
    \includegraphics[width=0.45\textwidth]{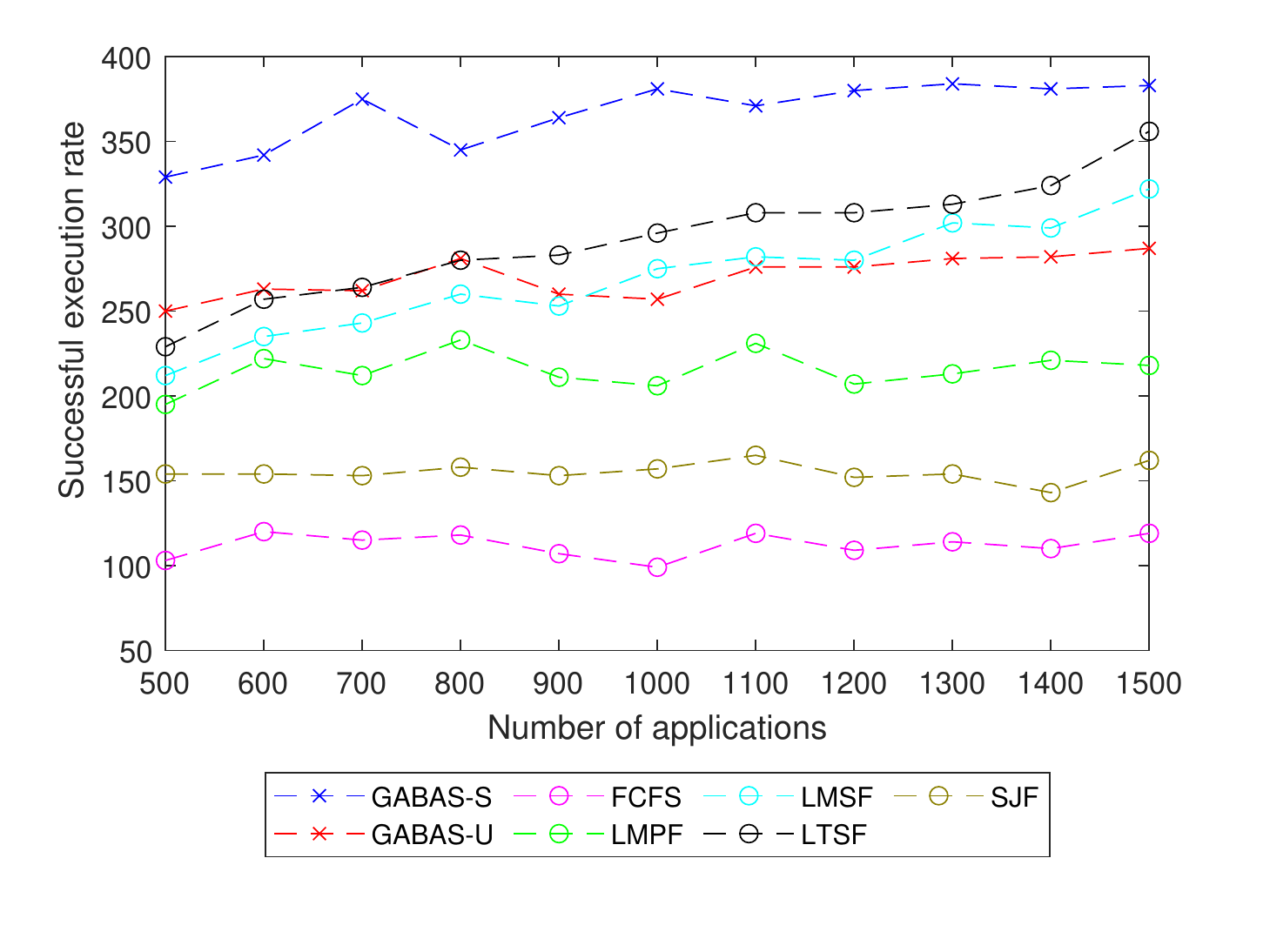}
    \caption{Comparison of algorithms in terms of successful execution rate in Scenario 1.}
    \label{scenario1suc}
\end{figure}

Figures \ref{scenario1wai}, \ref{scenario1res}, and \ref{scenario1suc} present the average waiting time, average turnaround time, and average successful execution rate in the first scenario, respectively. FCFS and SJF have the worst performance for all three metrics. GABAS-S is superior to all other algorithms. LTSF comes second, and LMSF comes third with a close performance to LTSF. In terms of average waiting and turnaround times, GABAS-U and LMPF have similar performance, but in terms of average successful execution rate, GABAS-U beats LMPF.

\begin{figure}[]
    \centering
    \includegraphics[width=0.45\textwidth]{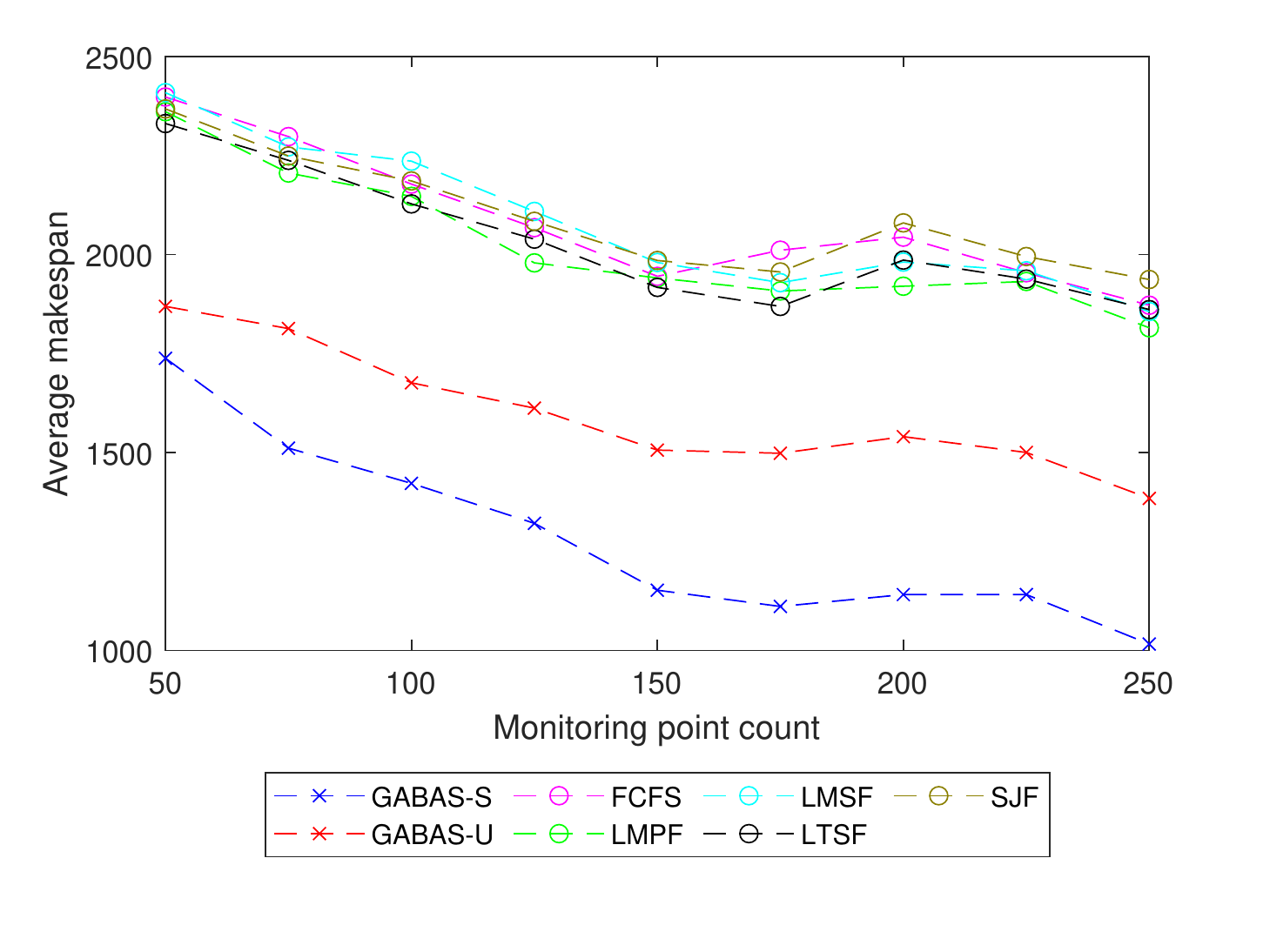}
    \caption{Comparison of algorithms in terms of average makespan in Scenario 2.}
    \label{scenario2avg}
\end{figure}

In Scenario 2, the simulations are done to observe the effect of the number of monitoring points on the performance of the algorithms. Figure \ref{scenario2avg} presents the average makespan for Scenario 2. GABAS-S is clearly superior to other algorithms, while GABAS-U comes second. LMPF again has the best performance among greedy algorithms. SJF generally has the worst performance of all. The performance gap between GABAS and others diminishes as the number of monitoring points in the region increases because it becomes easier to admit applications.

\begin{figure}[]
    \centering
    \includegraphics[width=0.45\textwidth]{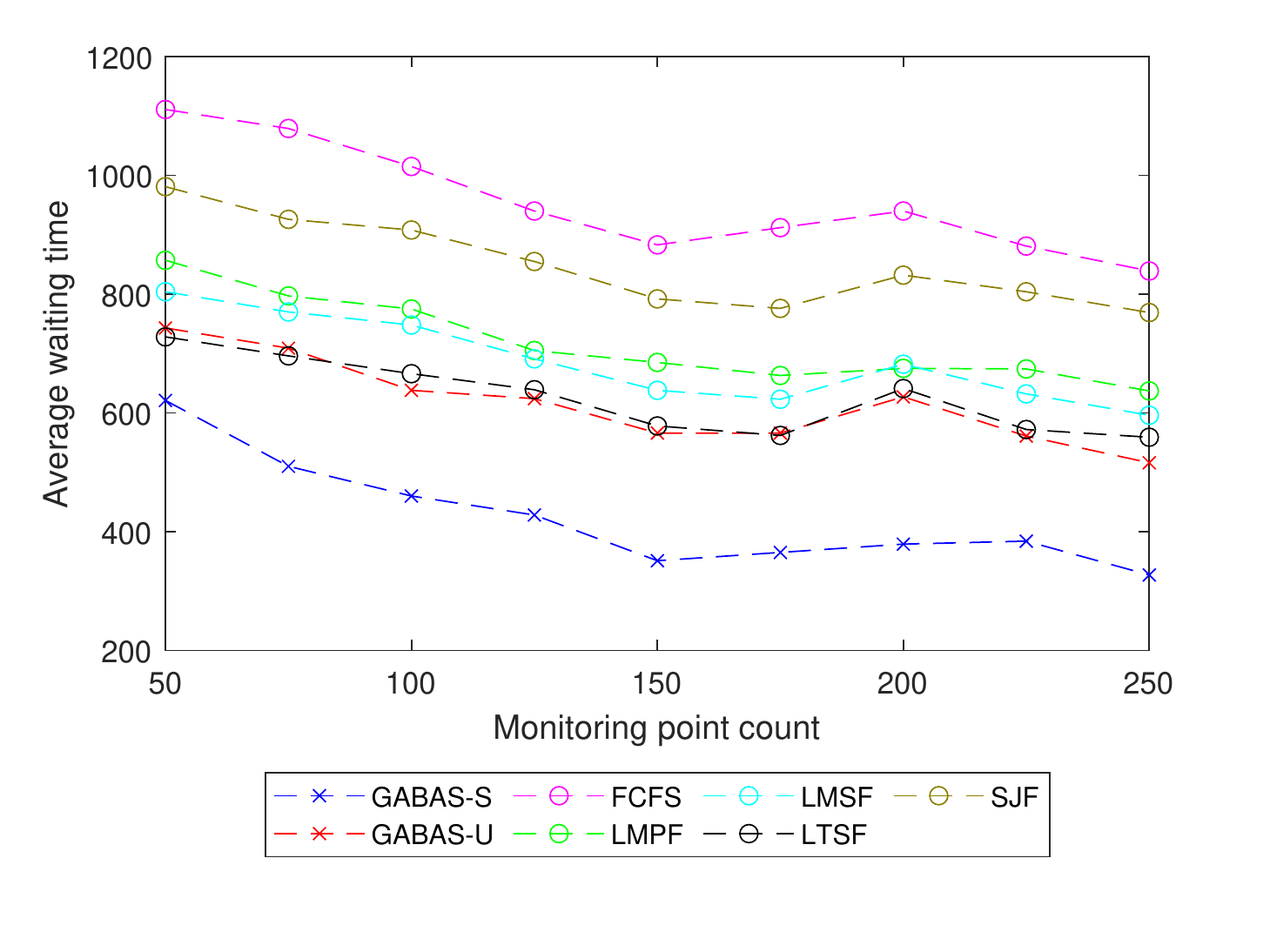}
    \caption{Comparison of algorithms in terms of average waiting time in Scenario 2.}
    \label{scenario2wai}
\end{figure}

\begin{figure}[]
    \centering
    \includegraphics[width=0.45\textwidth]{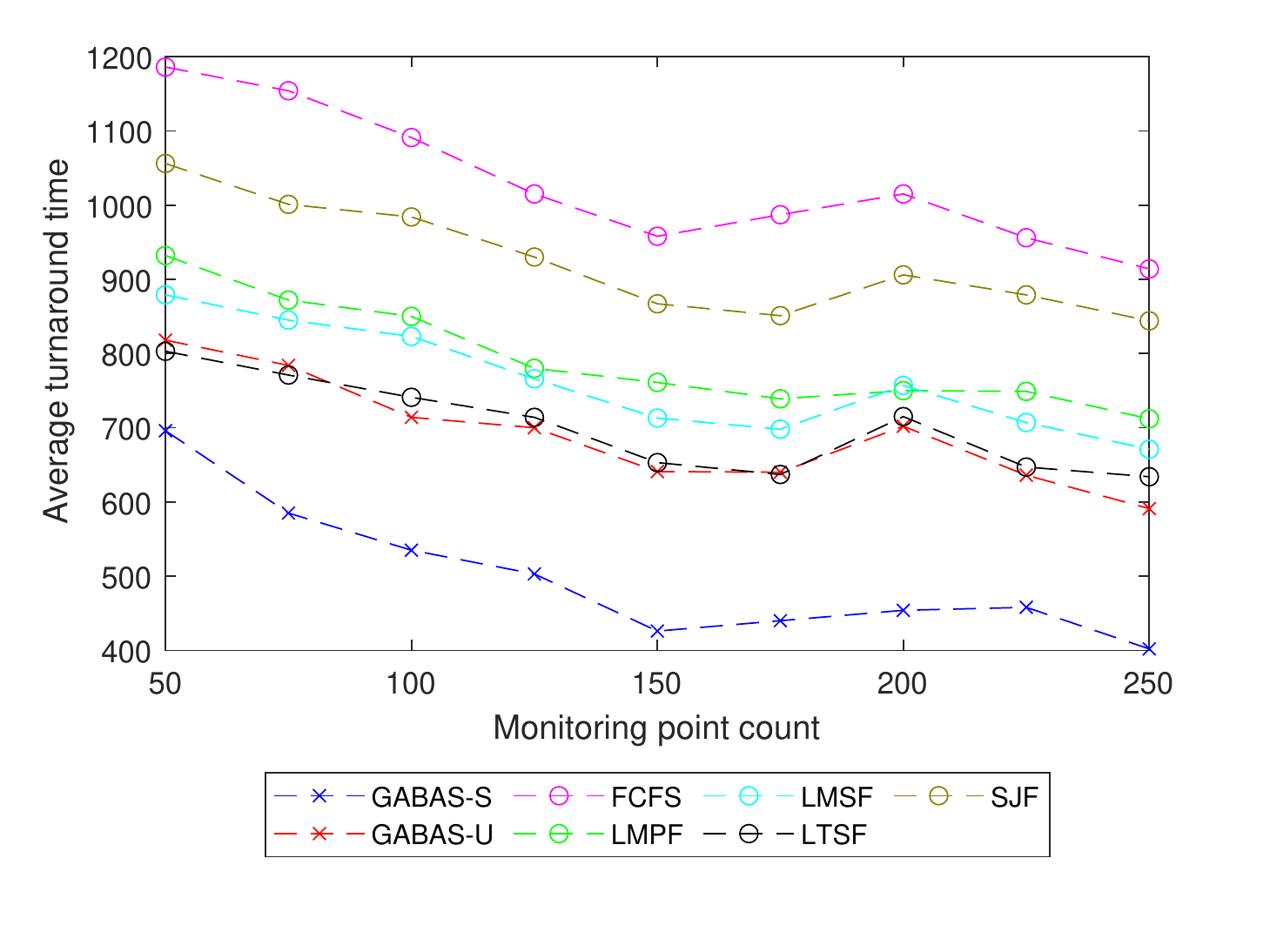}
    \caption{Comparison of algorithms in terms of average turnaround time in Scenario 2.}
    \label{scenario2res}
\end{figure}

\begin{figure}[]
    \centering
    \includegraphics[width=0.45\textwidth]{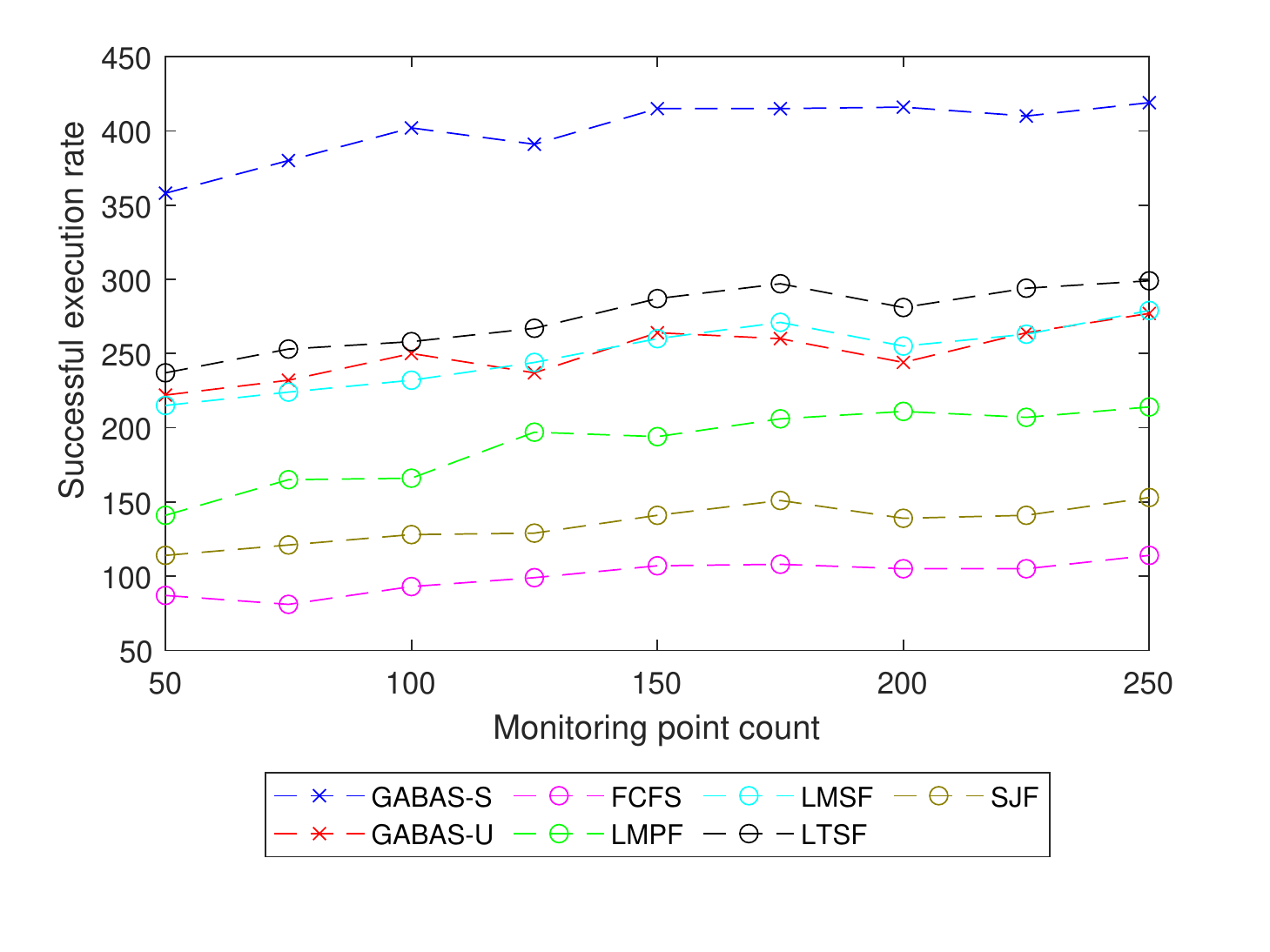}
    \caption{Comparison of algorithms in terms of successful execution rate in Scenario 2.}
    \label{scenario2suc}
\end{figure}

Average waiting time, average turnaround time, and average successful execution rate results of Scenario 2 are shown in Figures \ref{scenario2wai}, \ref{scenario2res}, and \ref{scenario2suc}, respectively. As in the previous scenario, FCFS and SJF have the worst performance among all algorithms, while GABAS-S has the best. The results for GABAS-U, LMSF, and LTSF are very close, but LMPF performs slightly worse among these algorithms. 

\begin{figure}[]
    \centering
    \includegraphics[width=0.45\textwidth]{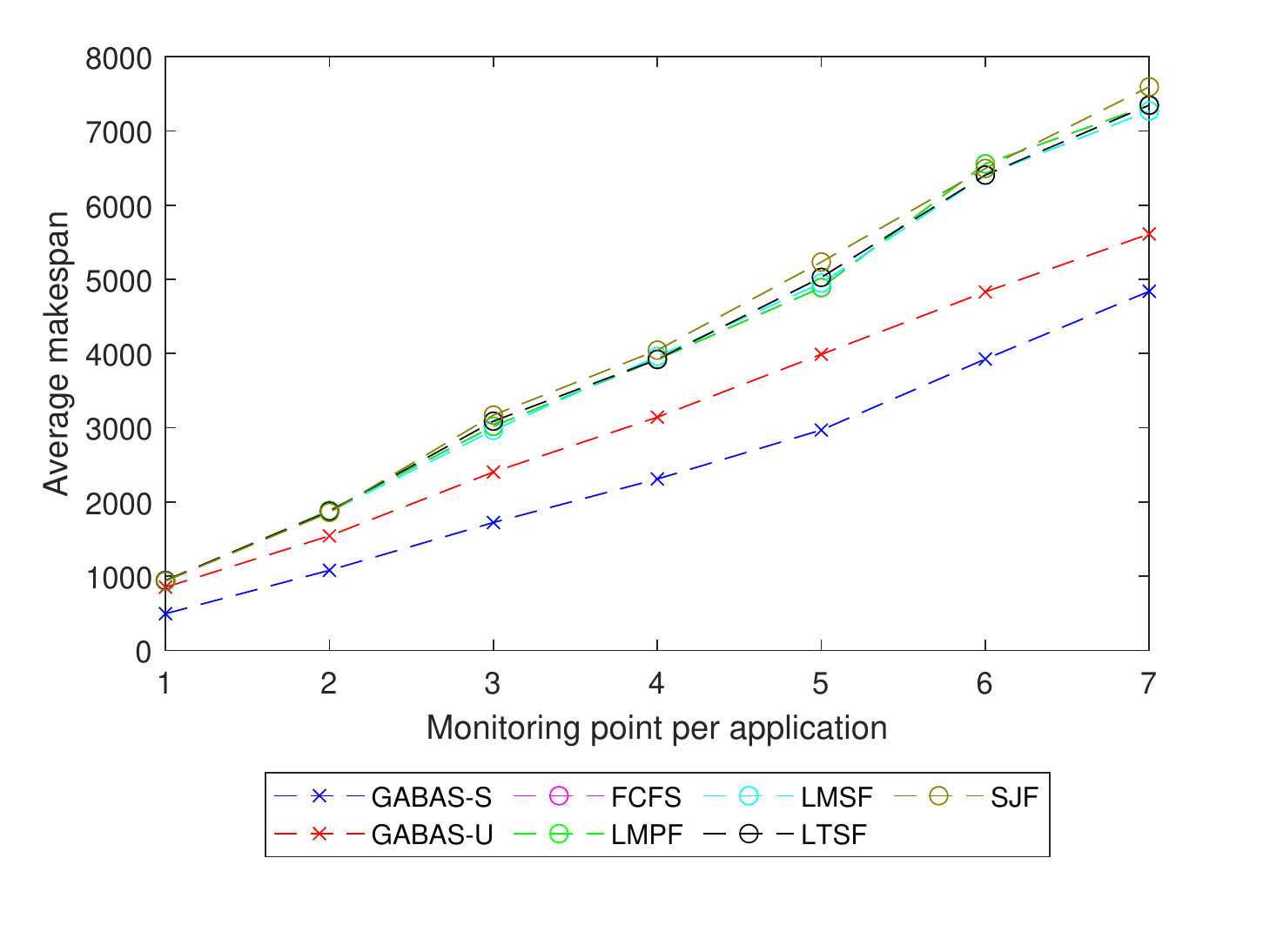}
    \caption{Comparison of algorithms in terms of average makespan in Scenario 3.}
    \label{scenario3avg}
\end{figure}

In Scenario 3, we aim to see the impact of the number of monitoring point requests per application on the algorithms' performance. Figure \ref{scenario3avg} presents the results of this scenario in terms of makespan. GABAS-S still has the best results. After five requests per application GABAS-U has the next best performance. Greedy approaches have very similar results. SJF is the slightly worst of all. In this scenario, LMPF behaves like FCFS since all applications have an equal number of requests. GABAS-S  performs better with a higher number of requests per application compared to the greedy algorithms. 

\begin{figure}[]
    \centering
    \includegraphics[width=0.45\textwidth]{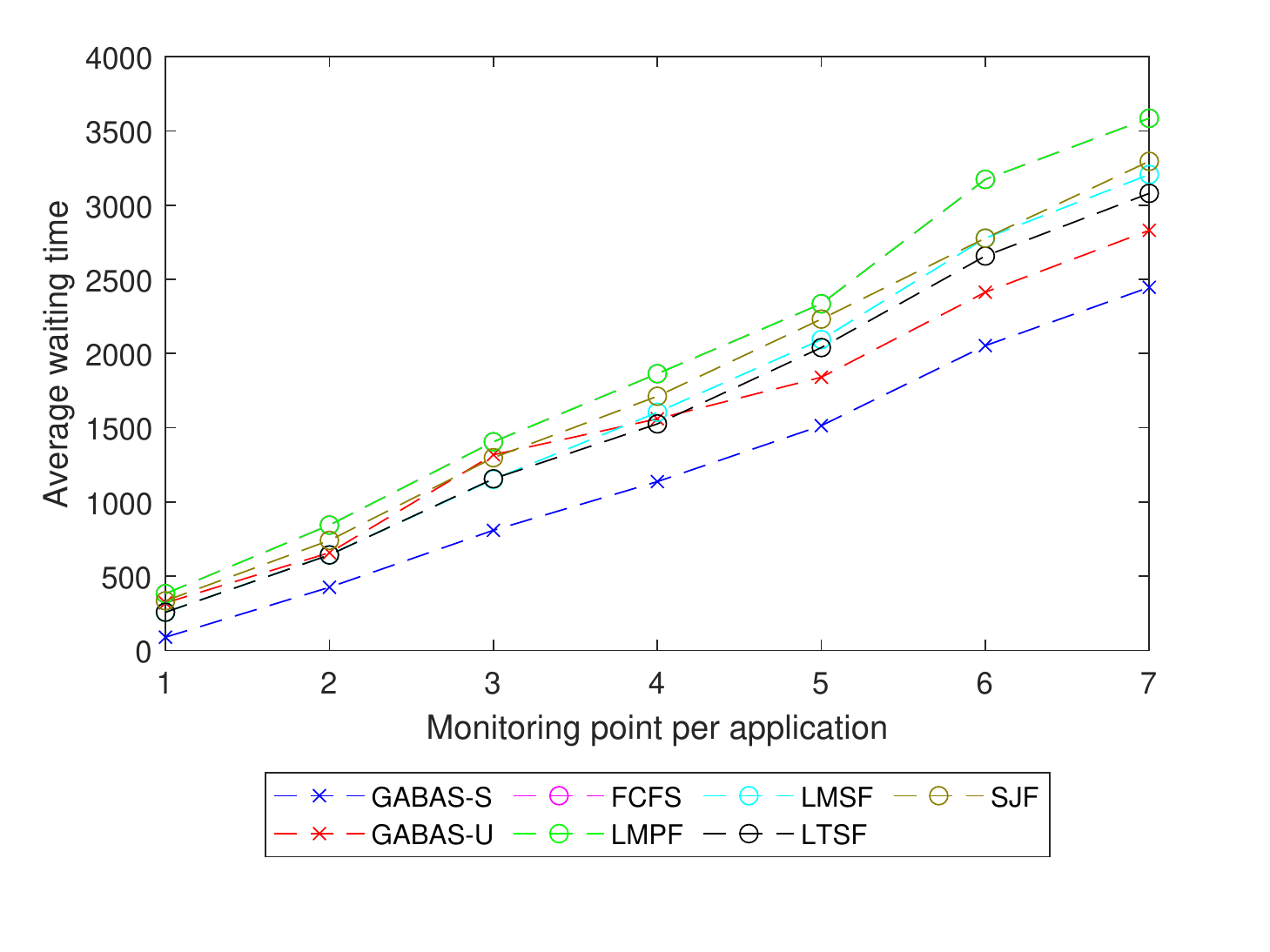}
    \caption{Comparison of algorithms in terms of average waiting time in Scenario 3.}
    \label{scenario3wai}
\end{figure}

\begin{figure}[]
    \centering
    \includegraphics[width=0.45\textwidth]{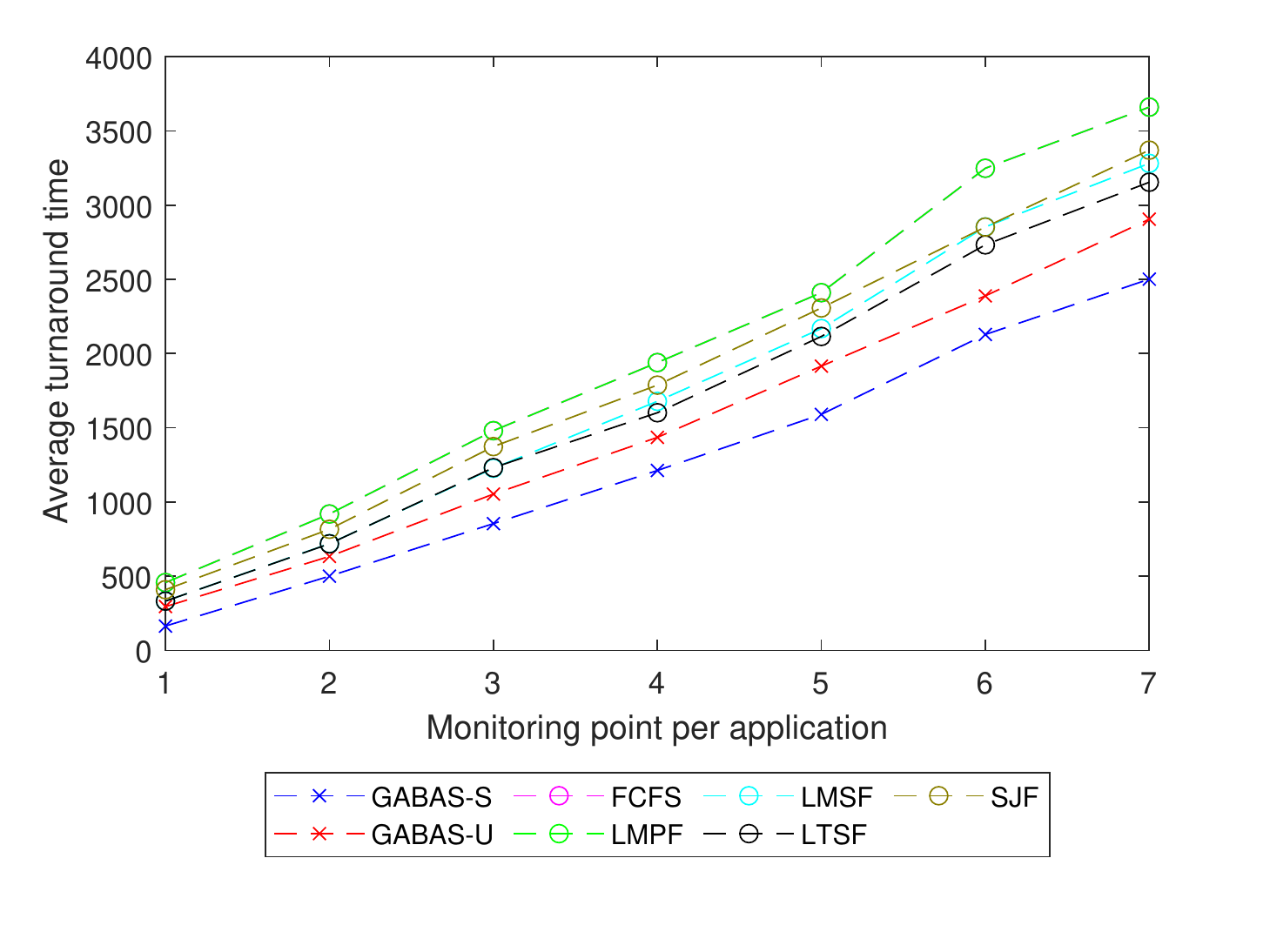}
    \caption{Comparison of algorithms in terms of average turnaround time in Scenario 3.}
    \label{scenario3res}
\end{figure}

\begin{figure}[]
    \centering
    \includegraphics[width=0.45\textwidth]{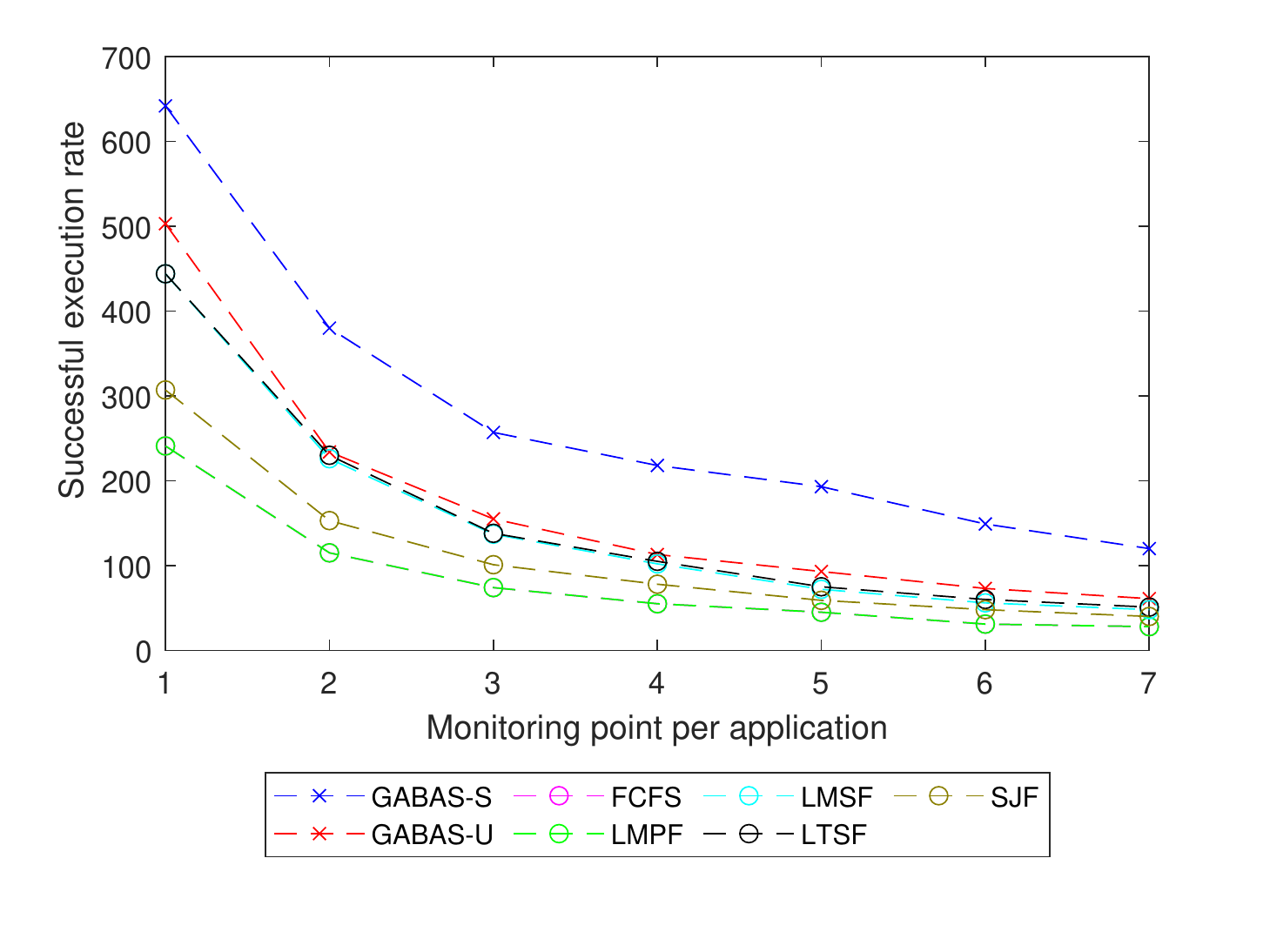}
    \caption{Comparison of algorithms in terms of successful execution rate in Scenario 3.}
    \label{scenario3suc}
\end{figure}

Figures \ref{scenario3wai}, \ref{scenario3res} and \ref{scenario3suc} display the results of average waiting time, turnaround time and successful execution rate, respectively. GABAS-S is the superior one among all. LMPF and FCFS have the worst performance for these criteria, while SJF is slightly better than them. GABAS-U has the second-best performance, while LMSF and LTSF produce the best results of all greedy algorithms. 

\begin{figure}[]
    \centering
    \includegraphics[width=0.45\textwidth]{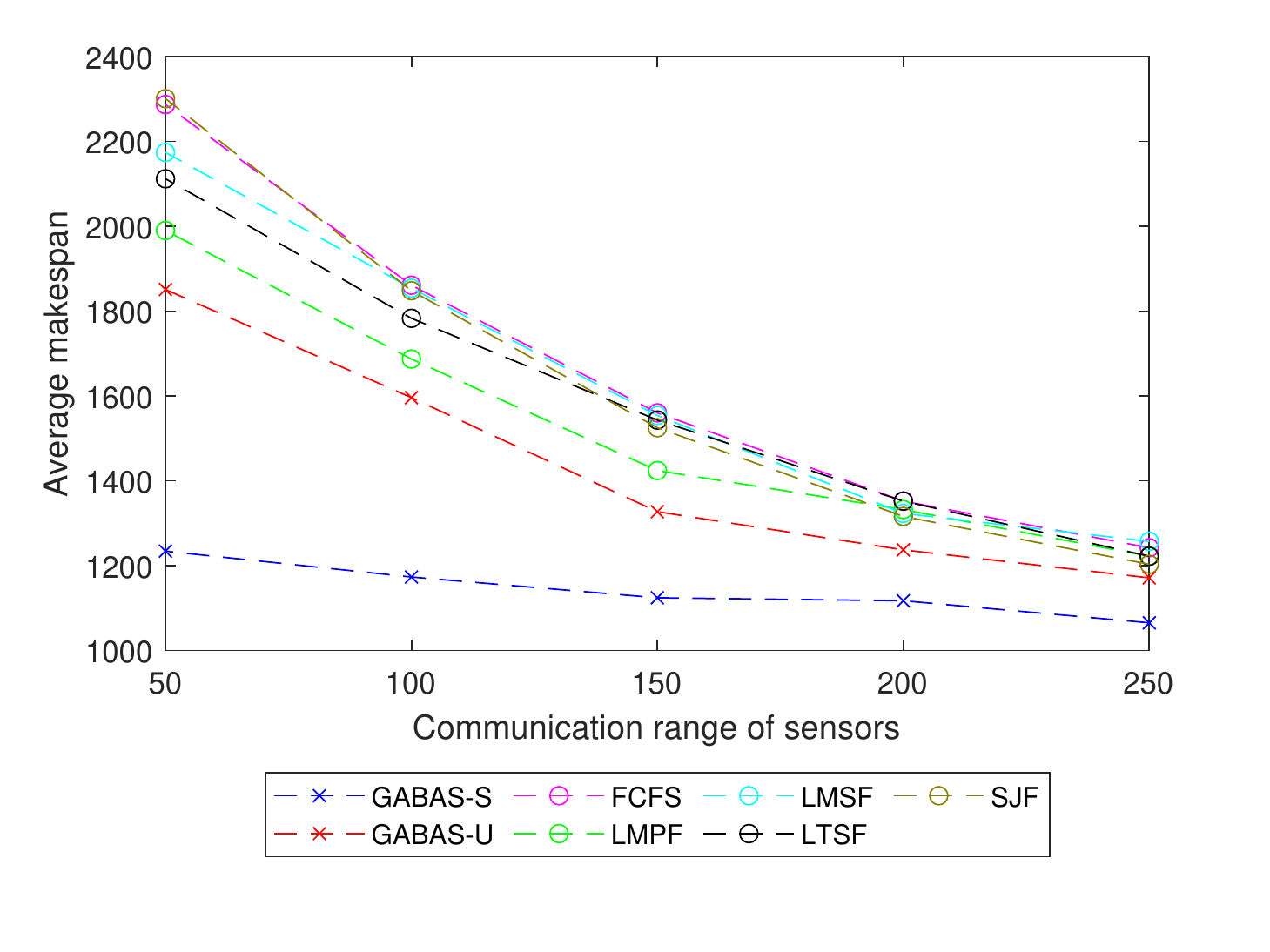}
    \caption{Comparison of algorithms in terms of average makespan in Scenario 4.}
    \label{scenario4avg}
\end{figure}

In Scenario 4, we evaluate the effect of change in the communication range of the sensor nodes on the results. Figure \ref{scenario4avg} presents the algorithms' makespan results. GABAS-S has the lowest makespan value, which makes it the best method among all compared algorithms. In terms of makespan, GABAS-U has the second-best performance. Greedy algorithms have similar results between 150-m and 250-m communication ranges. Between the 50-m and 150-m ranges, the performance of LMPF is distinguishable from the other four algorithms. SJF has the worst performance, especially with larger communication ranges.

\begin{figure}[]
    \centering
    \includegraphics[width=0.45\textwidth]{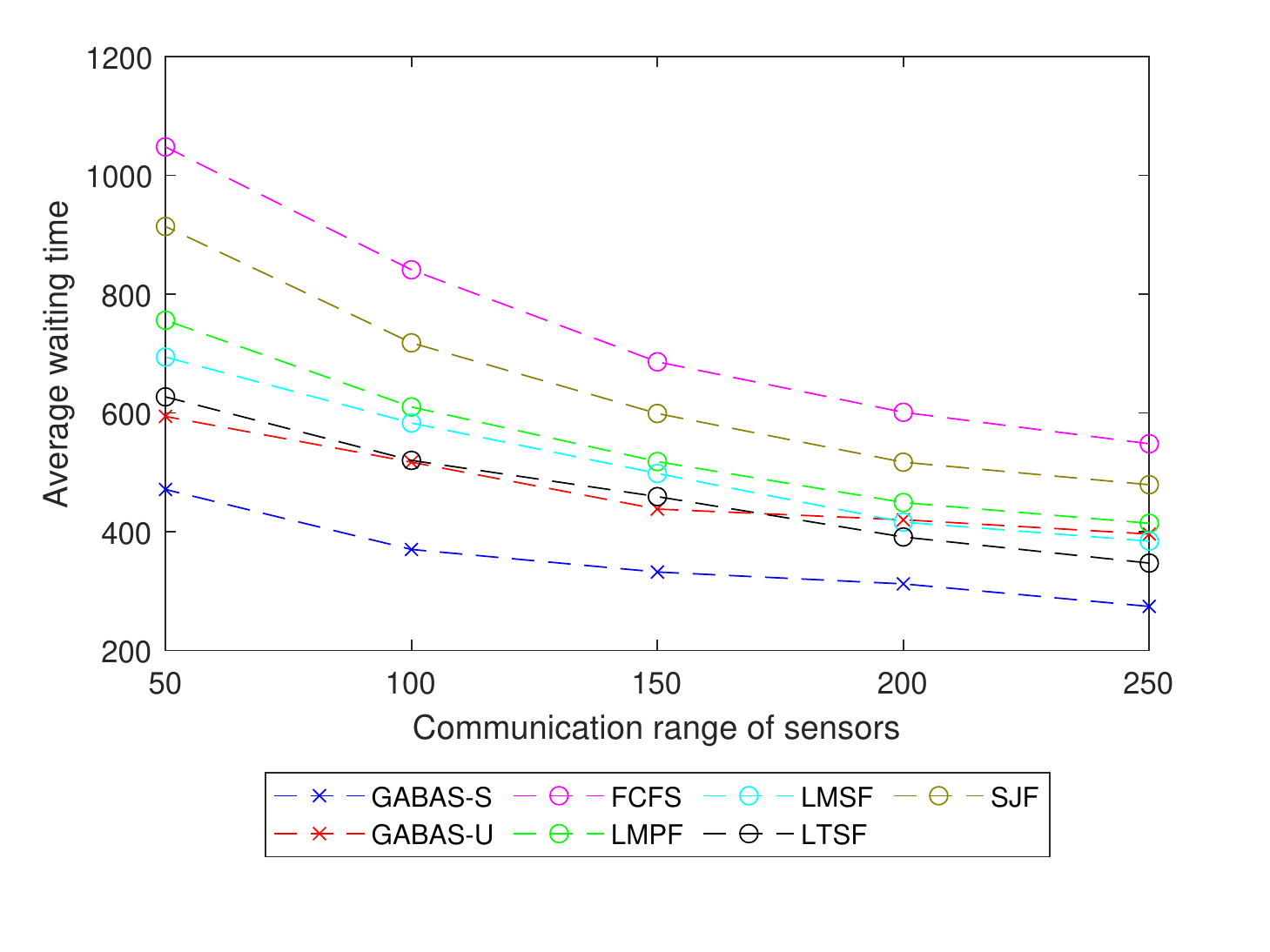}
    \caption{Comparison of algorithms in terms of average waiting time in Scenario 4.}
    \label{scenario4wai}
\end{figure}

\begin{figure}[]
    \centering
    \includegraphics[width=0.45\textwidth]{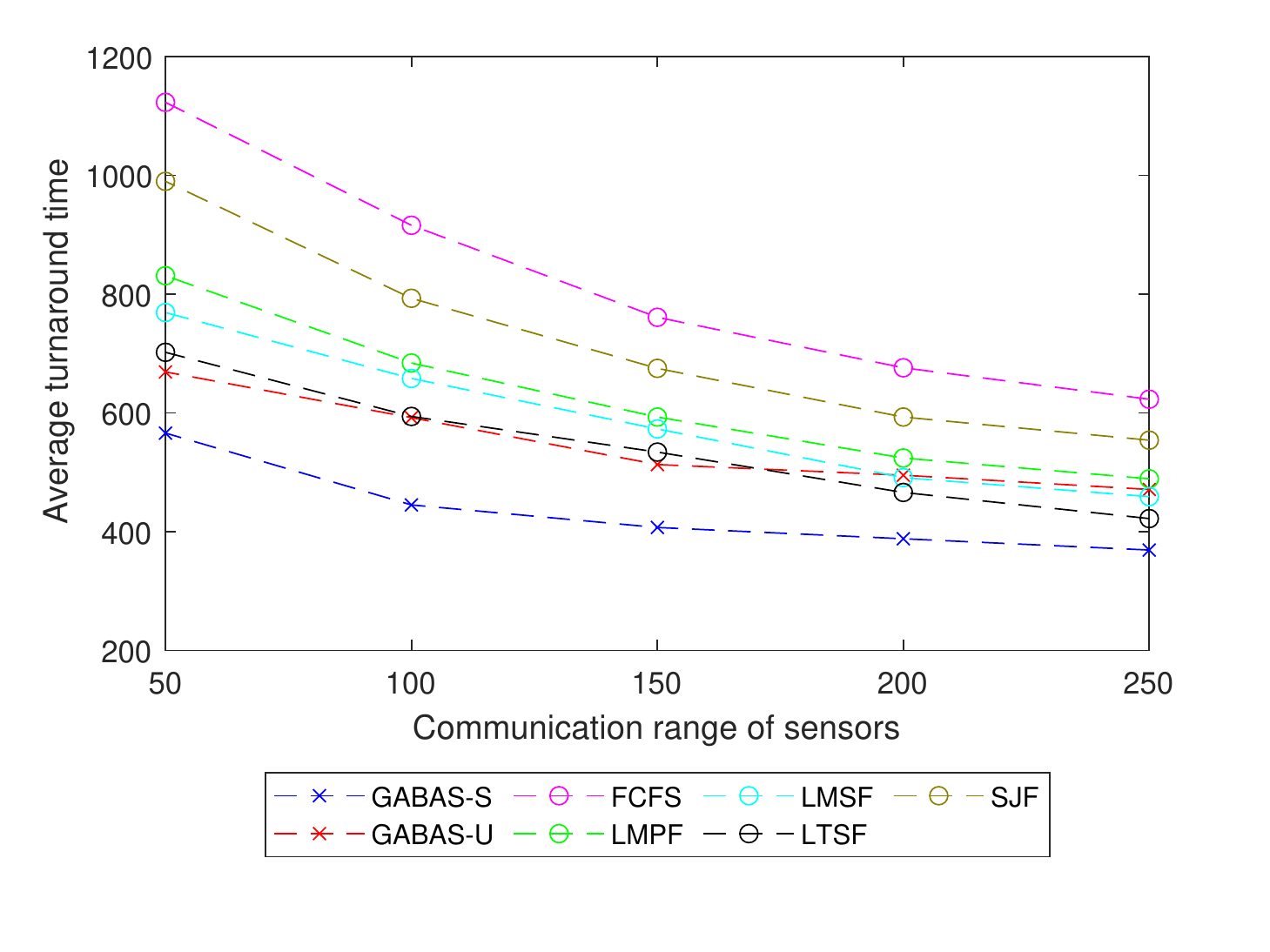}
    \caption{Comparison of algorithms in terms of average turnaround time in Scenario 4.}
    \label{scenario4res}
\end{figure}

\begin{figure}[]
    \centering
    \includegraphics[width=0.45\textwidth]{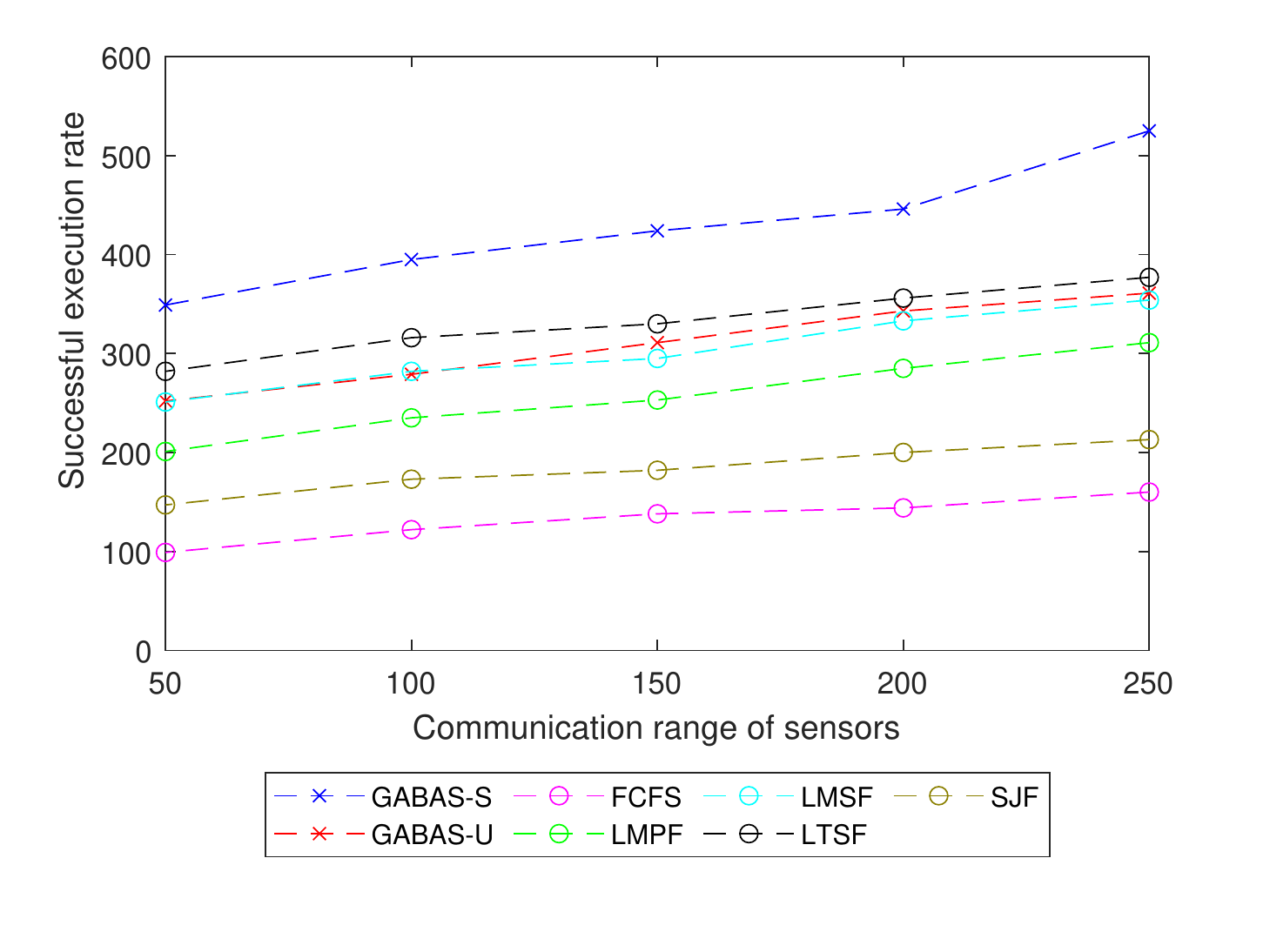}
    \caption{Comparison of algorithms in terms of successful execution rate in Scenario 4.}
    \label{scenario4suc}
\end{figure}

Average waiting and turnaround time for applications, and average successful execution rate are displayed in Figures \ref{scenario4wai}, \ref{scenario4res} and \ref{scenario4suc}, respectively. GABAS-S again has the superior performance. FCFS has the worst turnaround and waiting times as well as the least successful execution rate. GABAS-U, LTSF, and LMSF perform close to each other.

\begin{figure}[]
    \centering
    \includegraphics[width=0.45\textwidth]{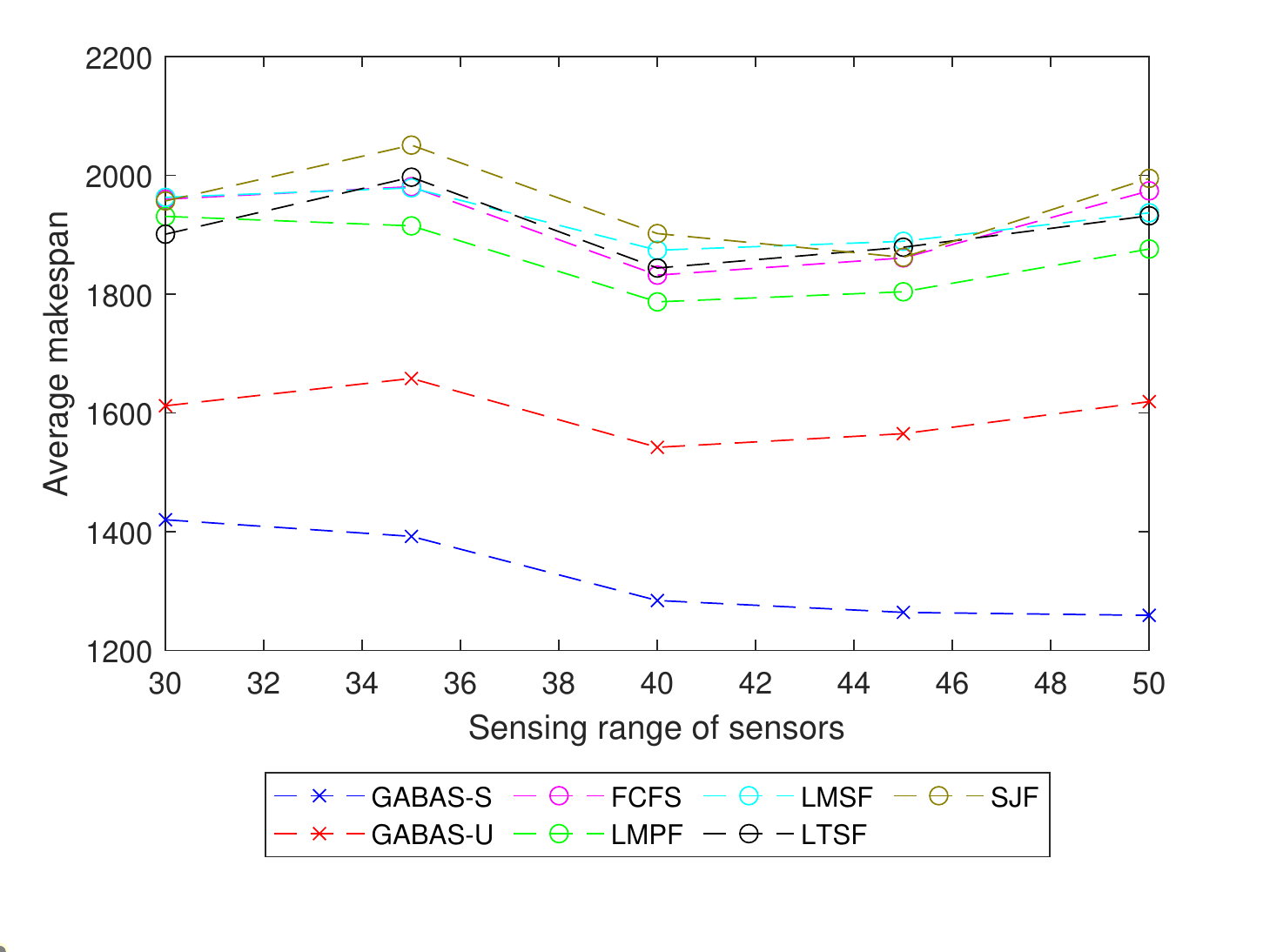}
    \caption{Comparison of algorithms in terms of average makespan in Scenario 5.}
    \label{scenario5avg}
\end{figure}

In Scenario 5, we investigate the impact of the sensing range of sensor nodes on the results. Makespan results for this scenario are shown in Figure~\ref{scenario5avg}. We can see that the sensing range does not affect the results drastically. With GABAS-S, the total execution time of applications is much smaller compared to other algorithms. GABAS-U produces the next best results, while the performance results of the greedy algorithms are close to each other. Again, LMPF has the best performance among all greedy algorithms in terms of makespan.

\begin{figure}[]
    \centering
    \includegraphics[width=0.45\textwidth]{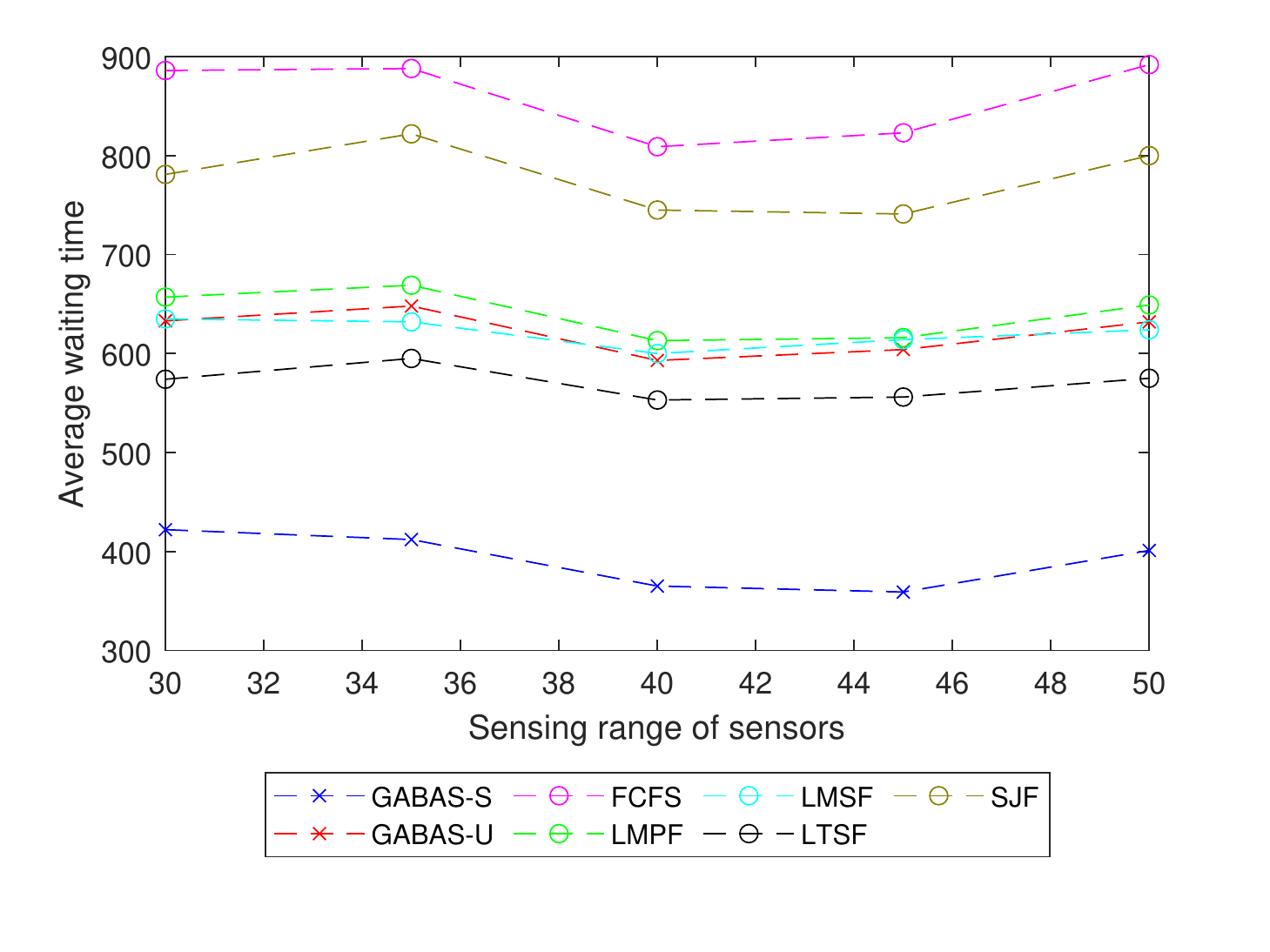}
    \caption{Comparison of algorithms in terms of average waiting time in Scenario 5.}
    \label{scenario5wai}
\end{figure}

\begin{figure}[]
    \centering
    \includegraphics[width=0.45\textwidth]{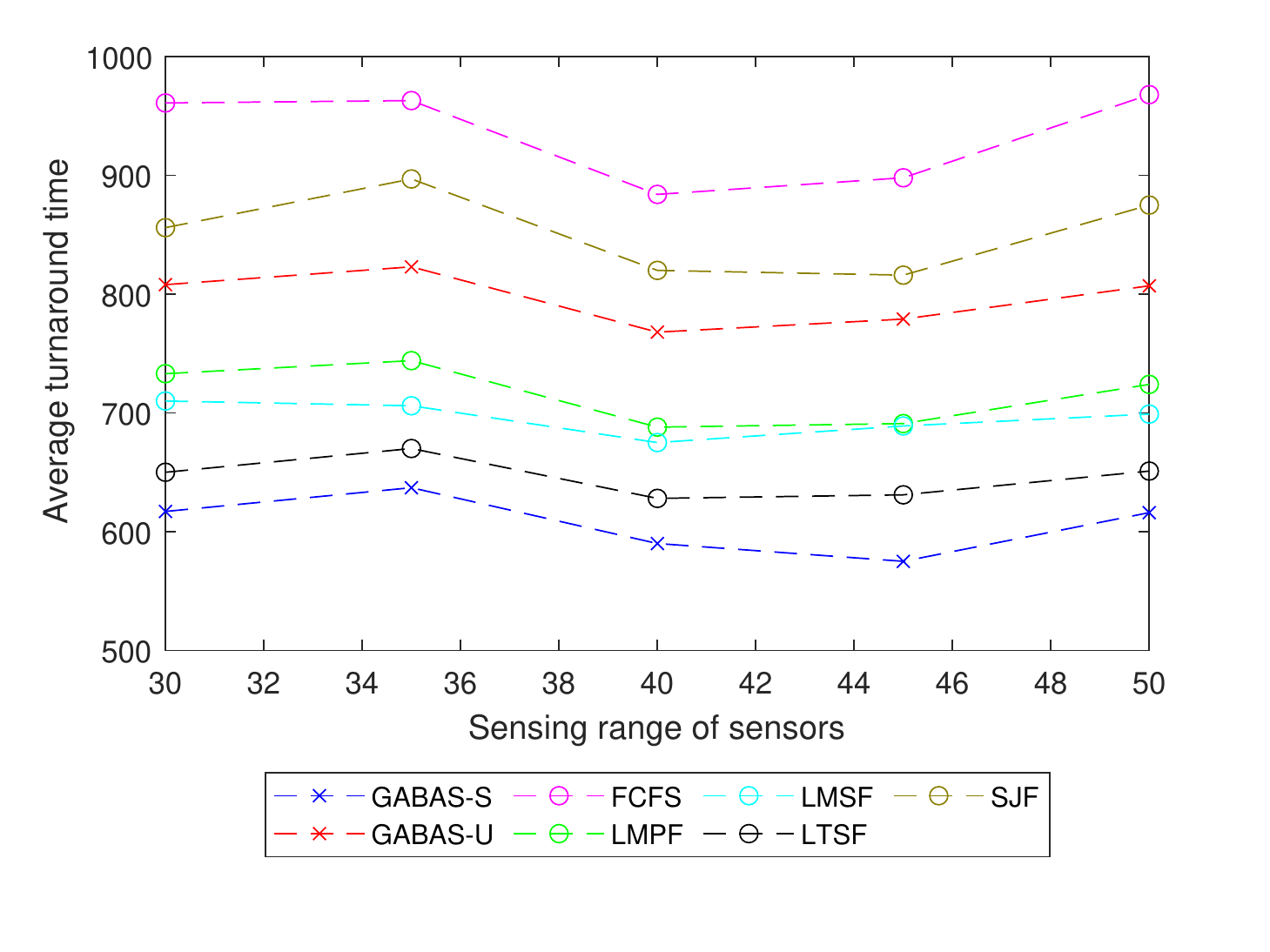}
    \caption{Comparison of algorithms in terms of average turnaround time in Scenario 5.}
    \label{scenario5res}
\end{figure}

\begin{figure}[]
    \centering
    \includegraphics[width=0.45\textwidth]{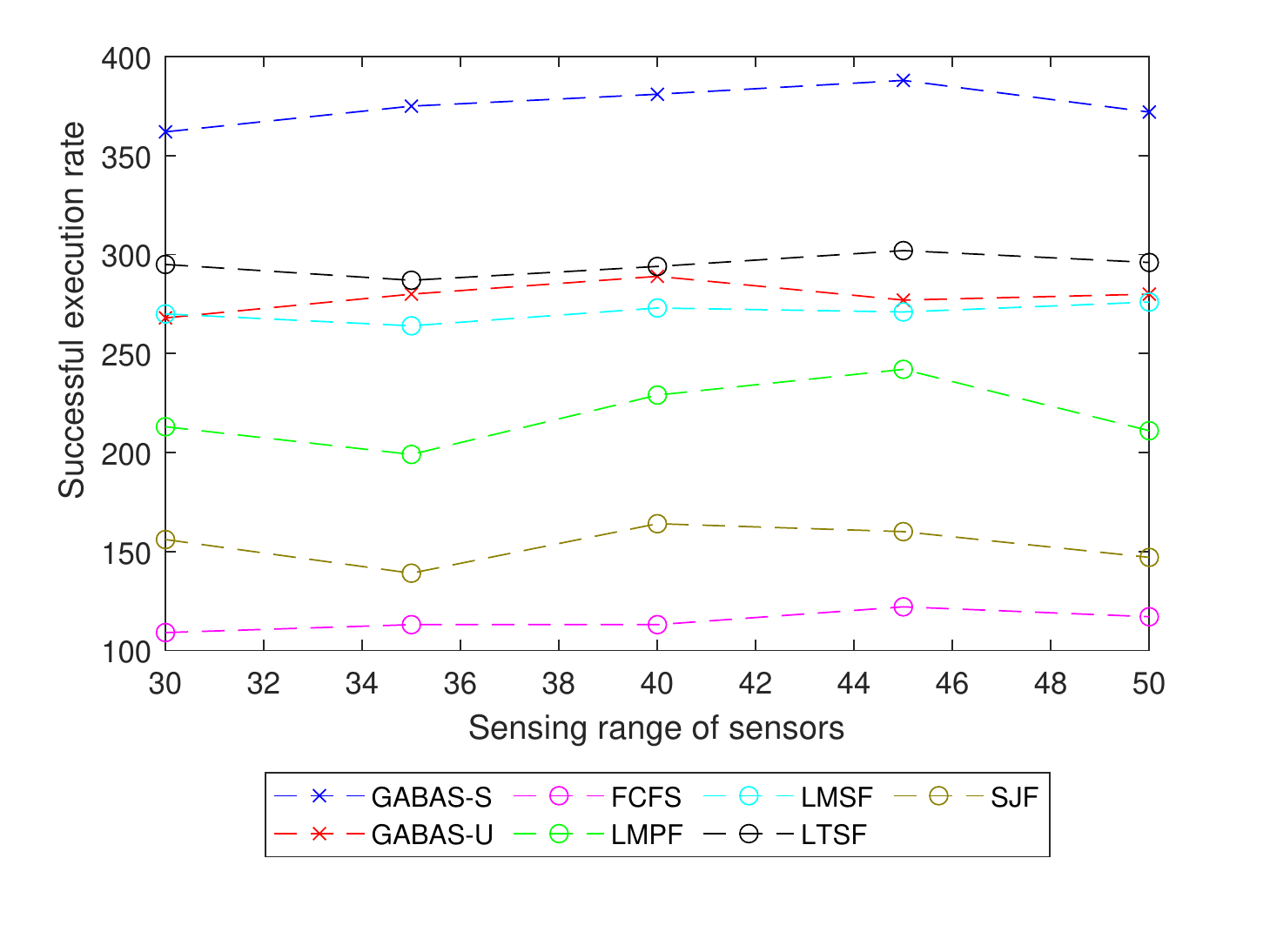}
    \caption{Comparison of algorithms in terms of successful execution rate in Scenario 5.}
    \label{scenario5suc}
\end{figure}

Average waiting time, turnaround time and successful execution rate results for this scenario are presented in Figures \ref{scenario5wai}, \ref{scenario5res} and \ref{scenario5suc}, respectively. Similarly, GABAS-S has the best performance, and FCFS has the worst. SJF is slightly better than FCFS but still behind the proposed greedy methods. LTSF comes second. GABAS-U, LMPF, and LMSF have close results in terms of waiting time; however, GABAS-U and LMPF perform slightly worse in terms of turnaround time and successful execution rate, respectively.

\begin{figure}[]
    \centering
    \includegraphics[width=0.45\textwidth]{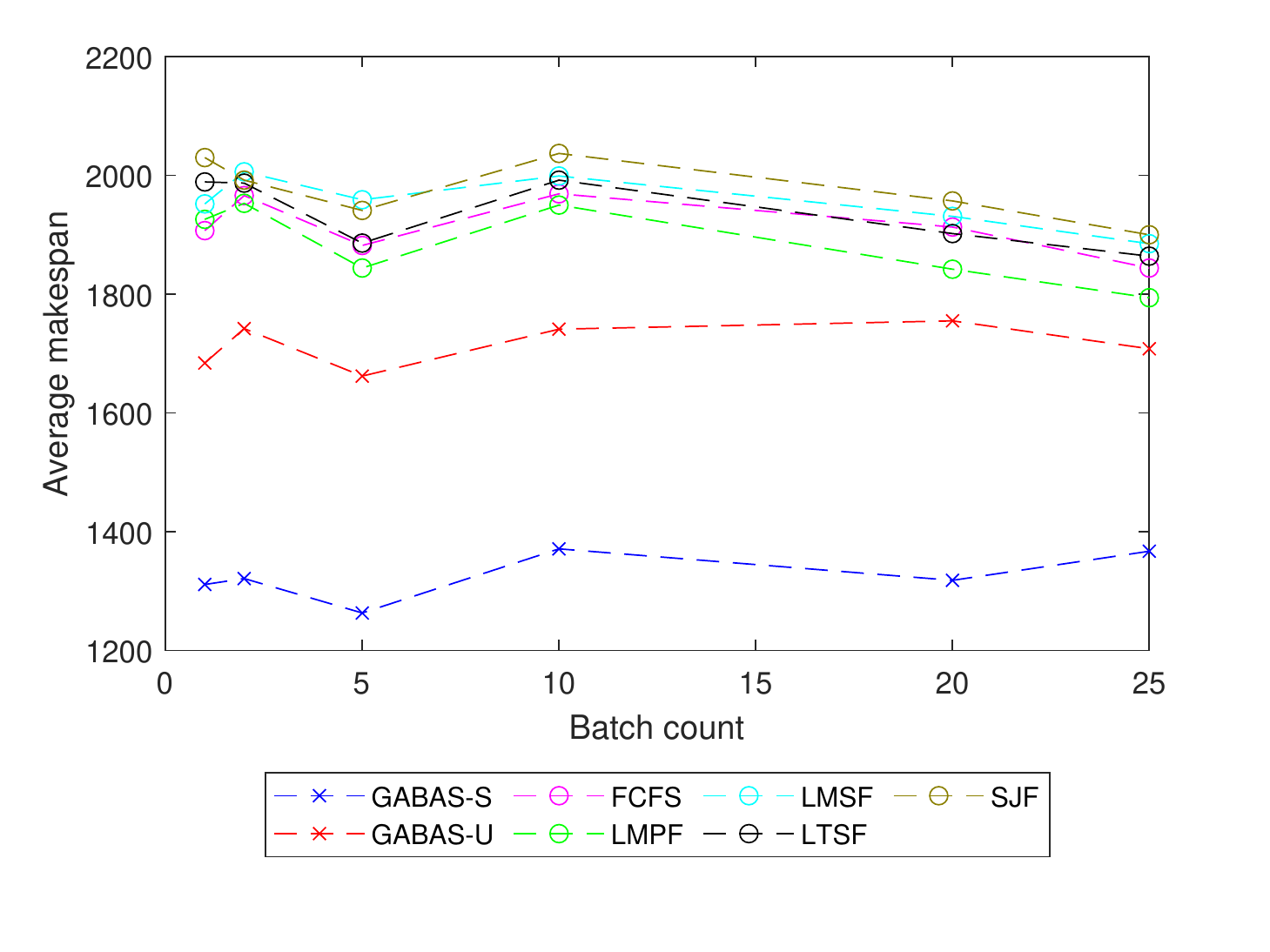}
    \caption{Comparison of algorithms in terms of average makespan in Scenario 6.}
    \label{scenario6avg}
\end{figure}

In Scenario 6, we observe how the number of batches affects the performance of the algorithms. Figure \ref{scenario6avg} shows the average makespan respectively in this scenario. In general, different batch counts do not affect much the performance of the algorithms. GABAS-S, again, has the best performance, which is followed by GABAS-U. Greedy methods have very similar performance, whereas LMPF is slightly better than the others.

\begin{figure}[]
    \centering
    \includegraphics[width=0.45\textwidth]{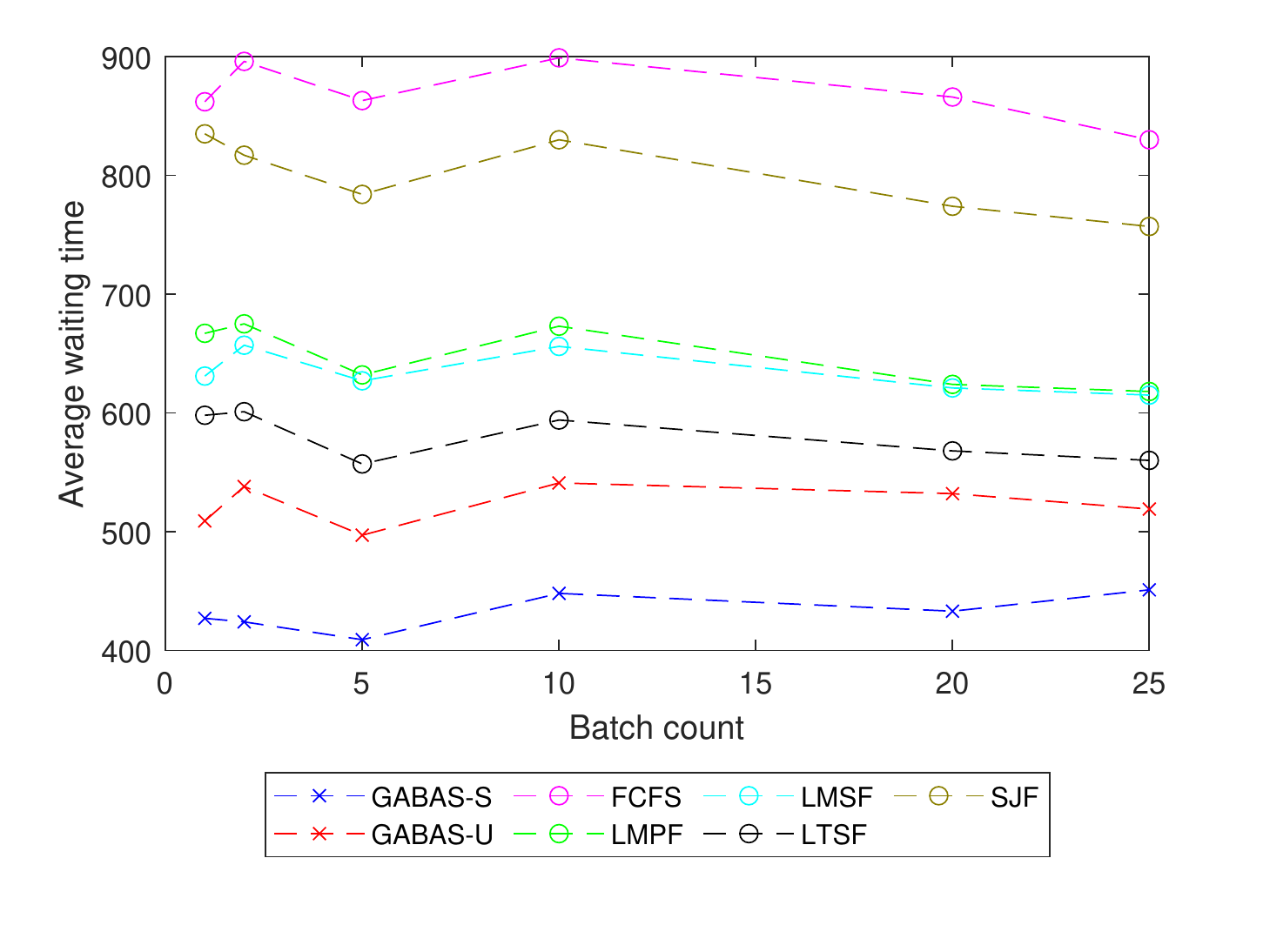}
    \caption{Comparison of algorithms in terms of average waiting time in Scenario 6.}
    \label{scenario6wai}
\end{figure}

\begin{figure}[]
    \centering
    \includegraphics[width=0.45\textwidth]{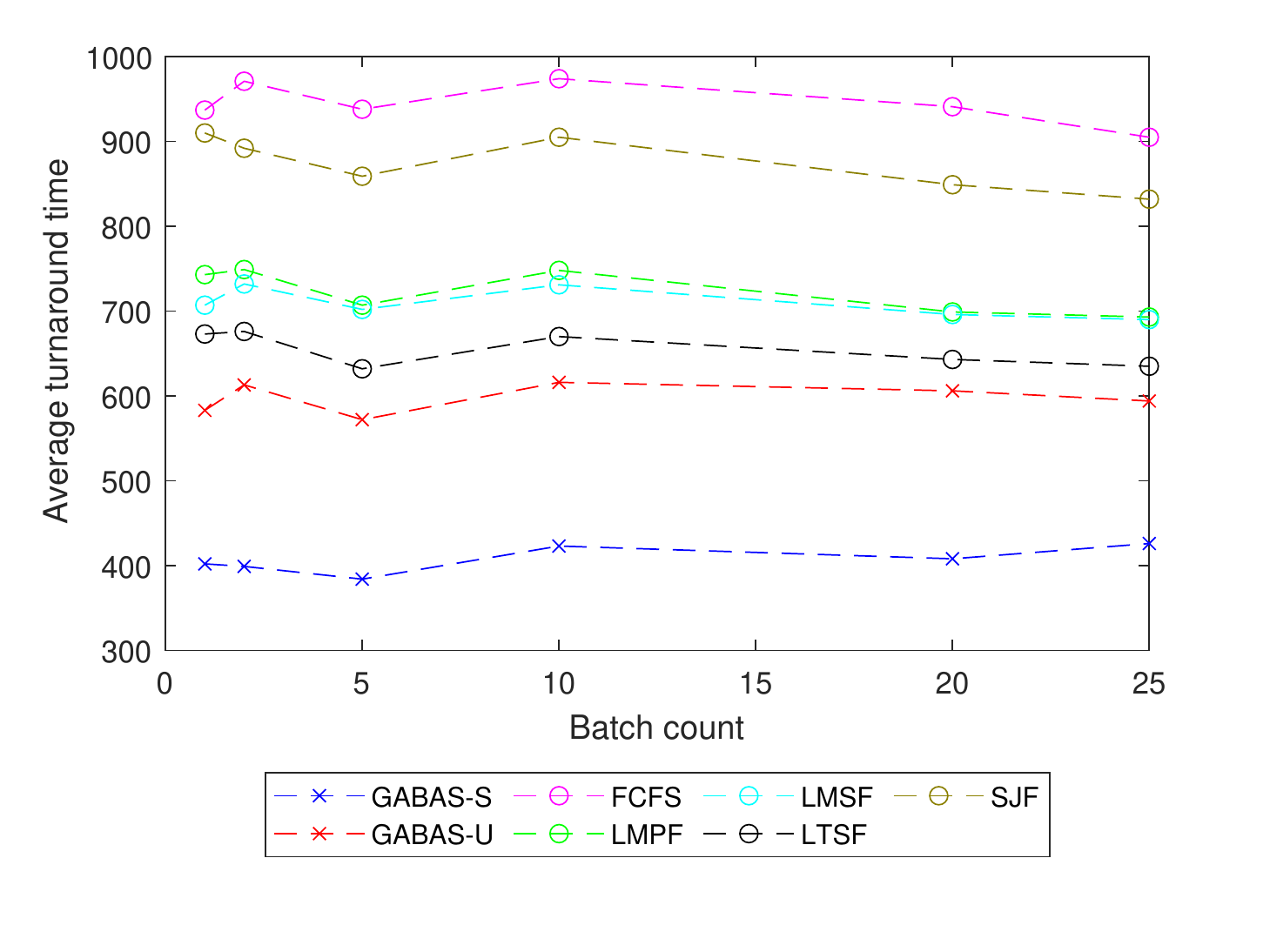}
    \caption{Comparison of algorithms in terms of average turnaround time in Scenario 6.}
    \label{scenario6res}
\end{figure}

\begin{figure}[]
    \centering
    \includegraphics[width=0.45\textwidth]{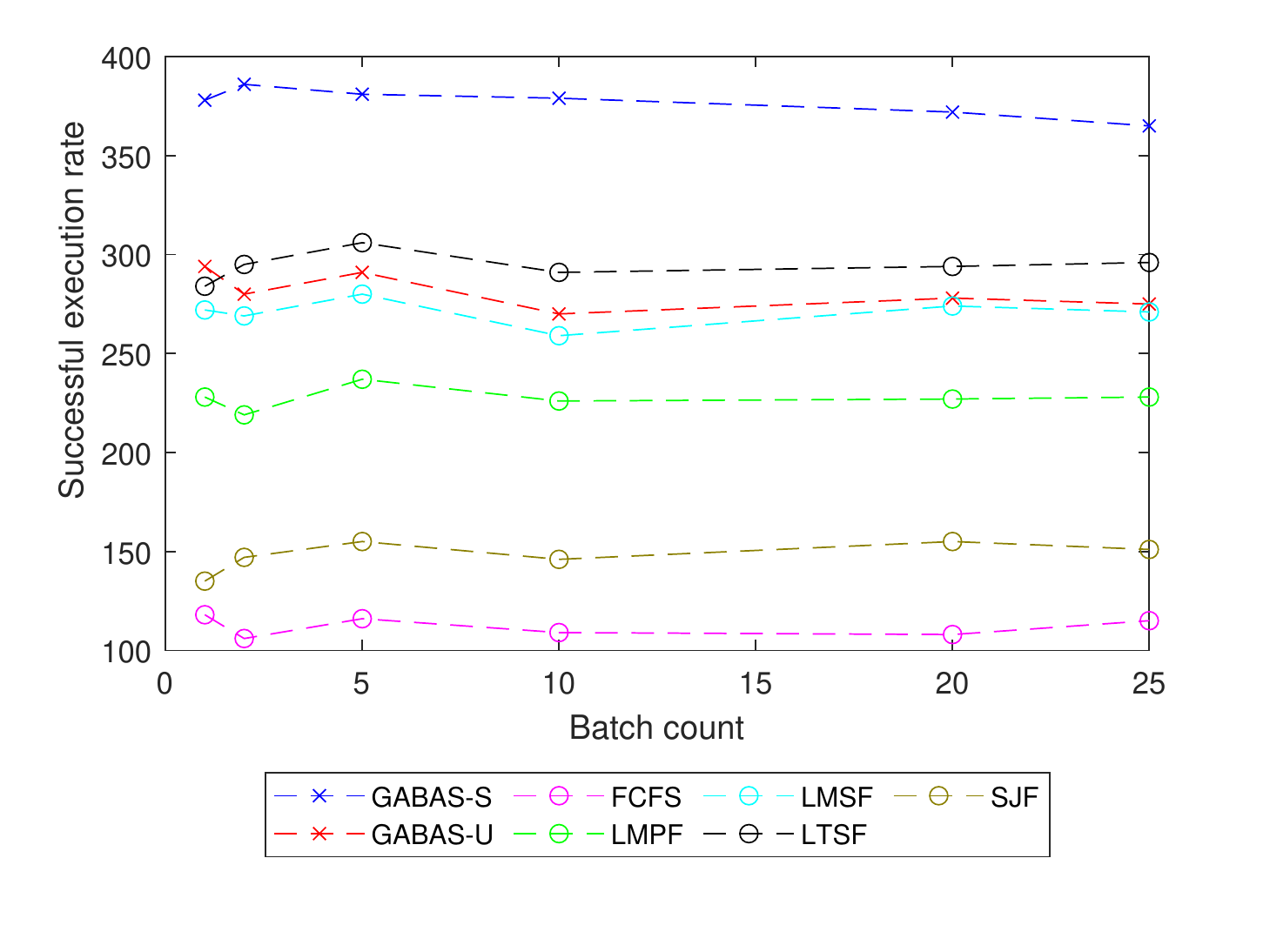}
    \caption{Comparison of algorithms in terms of successful execution rate in Scenario 6.}
    \label{scenario6suc}
\end{figure}

Regarding average waiting time, turnaround time, and successful execution rate, batch count again does not affect the results. Similar to the other scenarios, GABAS-S has the best performance. GABAS-U is slightly better than greedy methods. Among the greedy methods, LTSF performs the best which is followed by LMSF and LTSF. SJF and FCFS have the worst results. The waiting time, turnaround time, and successful execution rate results of this scenario are shown in Figures \ref{scenario6wai}, \ref{scenario6res}, and \ref{scenario6suc}, respectively.

We also measured the running times of the algorithms. A selected set of results are provided in Table~\ref{runtimes}. Our GABAS-S and GABAS-U algorithms are the slowest due to the nature of the genetic algorithms. GABAS-S is faster than GABAS-U since, with the shared-data approach, it is easier to place more applications at the same time; therefore, the total computation time is less compared to the unshared-data approach. Our greedy algorithms are much faster, as expected, compared to GABAS. Hence they are useful when fast decisions are required. FCFS is the fastest among all algorithms because it does not reorder the applications in the arrival queue.

\begin{table*}[]
\centering
\scriptsize
\caption{Running times of the algorithms in milliseconds.}
\begin{tabular}{|l|c|c|c|c|c|c|c|}
\hline
\multicolumn{1}{|c|}{Scenario} & GABAS-S & GABAS-U & LMPF & LMSF & LTSF & FCFS & SJF \\ \hline
Scenario1 \#A: 500 & 204 & 286 & 38 & 24 & 25 & 3 & 46 \\ \hline
Scenario1 \#A: 1000 & 352 & 524 & 105 & 63 & 70 & 2 & 90 \\ \hline
Scenario1 \#A: 1500 & 2588 & 3094 & 378 & 253 & 282 & 6 & 378 \\ \hline
Scenario2 \#MP: 50 & 377 & 899 & 93 & 67 & 71 & 3 & 103 \\ \hline
Scenario2 \#MP: 100 & 326 & 574 & 102 & 69 & 75 & 2 & 98 \\ \hline
Scenario2 \#MP: 250 & 229 & 289 & 73 & 51 & 64 & 1 & 92 \\ \hline
Scenario3 \#MP/A: 1 & 335 & 562 & 56 & 32 & 31 & 3 & 67 \\ \hline
Scenario3 \#MP/A: 3 & 1051 & 1508 & 362 & 256 & 245 & 3 & 413 \\ \hline
Scenario3 \#MP/A: 5 & 2254 & 5023 & 886 & 740 & 752 & 7 & 974 \\ \hline
\end{tabular}
\label{runtimes}
\end{table*}

In summary, from all our experimental results, we can draw the following conclusions:

\begin{itemize}

    \item GABAS-S is clearly superior to all other algorithms. It outperforms GABAS-U as well, and therefore we can conclude that the shared-data approach with multiplexed sensing is very effective in increasing the performance of the scheduling algorithms for various metrics.
    
    \item GABAS-U generally performs better than the greedy methods, even if it uses the unshared-data approach, while all the greedy methods use the shared-data approach. Even in the experiments measuring waiting time, turnaround time, and successful execution rate, GABAS-U has a better performance compared to greedy algorithms, especially when network resources are scarcer, even though it does not target these metrics directly. This result shows that using meta-heuristic algorithms is very effective in the application scheduling problem for WSNs.
    
    \item In terms of makespan, LMPF is the best greedy algorithm among the compared algorithms.
    
    \item In terms of waiting time, turnaround time, and successful execution rate, LMSF and LTSF have the best performance among greedy and standard algorithms.
    
    \item All proposed algorithms perform better than the standard FCFS and SJF algorithms.
    
\end{itemize}

\section{Conclusion}

In this paper, we studied the application scheduling problem in wireless sensor networks. First, we proposed a shared-data approach that provides an opportunity to perform multiplexed sensing of monitoring points for multiple applications that can share data, and in this way to reduce sensing and communication resource usage. Then, we proposed a genetic algorithm called GABAS, for scheduling applications effectively. We also proposed three greedy algorithms, LMPF, LMSF, and LTSF, that can be used for scenarios where fast decisions are needed. All our proposed algorithms decide both on the assignments of sensor nodes and base stations to monitoring points and the admission order of the waiting applications. We compared our proposed algorithms with each other and the well-known task scheduling algorithms, First Come First Served and Shortest Job First, in terms of makespan, waiting time, turnaround time, and successful execution rate by performing extensive simulation experiments. We observed that GABAS outperforms all other algorithms in all comparison metrics. Among other algorithms, LMPF provides the best results in terms of makespan, and LMSF and LTSF provide the best performance in terms of waiting time, turnaround time, and successful execution rate.

\appendices
\section{Reduction of Application Scheduling to Multiway Number Partitioning}

\subsection{Multiway Number Partitioning}

Multiway number partitioning (MNP) is the problem of partitioning a multi-set of numbers into \textit{k} different subsets in a way that sums of numbers in each subset are as similar as possible. It is a generalized version of the partitioning problem where $k=2$. Partitioning is proven to be NP-hard~\cite{karp1972reducibility}.

\subsection{Reduction to Application Scheduling}

We assume that there are \textit{n} applications waiting to be deployed. Each application requires a single monitoring point to be sensed with a unit sensing rate. Applications need to be deployed to the network for a certain amount of time which is denoted by $t_{j}$ for application \textit{j}. There are \textit{k} sensor nodes and each sensor node is connected to a single base station. Each sensor node and the base station have unit sensing and processing capacity, respectively. Any sensor node to base station connection has unit bandwidth. Both transmission and processing coefficients are equal to 1. Each monitoring point is required to be sensed by a single application.

In MNP, we have a set $S$ of $n$ numbers $a_{1}, a_{2},.., a_{n}$. The set is to be partitioned into \textit{k} subsets such that the maximum subset-sum is minimized. Transformation is done as follows:

Each number in the MNP set is the running time requirement of an application. In other words, $a_{j} = t_{j}$, where $t_{j}$ is the running time requirement of application $j$. Each partition corresponds to a sensor node and base station pair that will take part in sensing and processing the data of the applications assigned to them. If $a_{j}$ is in partition \textit{i}, then the sensing and processing requirement of the application \textit{j} for a monitoring point is handled by the sensor node and base station pair \textit{i}. The sum of numbers assigned to a partition represents the amount of time during which the corresponding sensor node and base station pair will be active, i.e., sensing and processing for the applications assigned to them. The maximum sum among the sums for all partitions is equal to the maximum active time of a sensor node and base station pair, which is equal to the finish time of the last application. Therefore, minimizing the maximum sum is equal to minimizing $tf_{max}$ in application scheduling.

\bibliographystyle{IEEEtran}
\bibliography{mybib.bib}

\vfill

\end{document}